\documentclass[aps, prd, showkeys, showpacs ]{revtex4}

\usepackage{amsmath,amsfonts,amscd,amssymb,epsf, multirow}
\usepackage[small]{caption}

\newcommand{\pa}{\partial}
\newcommand{\vep}{\varepsilon}
\begin{document}
\title{Scalar cylinder-plate and cylinder-cylinder Casimir interaction  in higher dimensional spacetime }
\author{Lee-Peng Teo}
 \email{LeePeng.Teo@nottingham.edu.my}
 \affiliation{Department of Applied Mathematics, Faculty of Engineering, University of Nottingham Malaysia Campus, Jalan Broga, 43500, Semenyih, Selangor Darul Ehsan, Malaysia.}
\begin{abstract}
We study the cylinder-plate and the cylinder-cylinder Casimir interaction in the $(D+1)$-dimensional Minkowski spacetime due to the vacuum fluctuations of massless scalar fields. Different combinations of Dirichlet (D) and Neumann (N) boundary conditions are imposed on the two interacting objects. For the cylinder-cylinder interaction, we consider the case where one cylinder is inside the other, and the case where the two cylinders are outside each other. By computing the transition matrices of the objects and the translation matrices that relate different coordinate systems,   the explicit formulas for the Casimir interaction energies are derived. From these formulas, we compute the large separation and small separation asymptotic behaviors of the Casimir interaction. For  the cylinder-plate interaction with $R\ll L$, where $R$ is the radius of the cylinder and $L$ is the distance from the center of the cylinder to the plate,  the order of decay of the Casimir interaction only depends on the boundary conditions imposed on the cylinder. The orders are $L^{-D+1}/\ln (L)$ and $L^{-D-1}/\ln L$ respectively for Dirichlet and Neumann boundary conditions on the cylinder. For two cylinders with radii $R_1$ and $R_2$ lying parallelly outside each other, the orders of decay of the Casimir interaction energies when $R_1+R_2\ll L$ are $L^{-D+1}/(\ln L)^2$, $L^{-D-1}/\ln L$ and $L^{-D-3}$ respectively for DD, DN/ND and NN boundary conditions, where $L$ is the distance between the centers of the cylinders. The more interesting and important characteristic of Casimir interaction appears at small separation. Using perturbation technique, we compute the small separation asymptotic expansions of the Casimir interaction energies up to the next-to-leading order terms. The leading terms coincide with the respective results obtained using proximity force approximation, which is of order $d^{-D+1/2}$, where $d$ is the distance between the two objects. The  results on the next-to-leading order terms are more interesting and important. We find some universal behaviors. It is also noticed that for the case of Dirichlet-Dirichlet cylinder-plate interaction, the next-to-leading order term agrees with that obtained using derivative expansion. Hence, based on our results on other boundary conditions and on the cylinder-cylinder interaction, we postulate a formula for the derivative expansion to expand the Casimir interaction energy up to the next-to-leading order terms for DD, DN, ND and NN boundary conditions, for the interaction between two curved surfaces in $(D+1)$-dimensional Minkowski spacetime. It is found that the postulate agrees with our previous results on the sphere-sphere interactions except when $D=4$.
\end{abstract}

 \maketitle

\section{Introduction}
In the pioneering work \cite{1}, Casimir proposed the existence of a force between  two parallel perfectly conducting plates due to the vacuum fluctuations of electromagnetic fields. This gives rise to the concept of vacuum energy, which was referred to as Casimir energy.  In subsequent years, Casimir effect has been generalized to any quantum fields, and it is a purely quantum effect. The idea of this Casimir energy is quite simple. The ground state energy of a quantum harmonic oscillator is not zero, but it is equal to $\hbar\omega/2$, where $\omega$ is the frequency. A quantum field can be considered as the superposition of an infinite number of quantum harmonic oscillators, each with a different ground state energy. Casimir defined the Casimir energy as the sum of the ground state energies:
\begin{equation}
E_{\text{Cas}}=\sum \frac{\hbar\omega}{2}.
\end{equation} This sum is divergent and regularization is required. However, in the existence of two objects (boundaries), one can obtain a \emph{finite} Casimir interaction energy after subtracting away the Casimir self energies of each of the objects.

The Casimir self energy of an object is of its own interest and it has been under active investigation  (see \cite{2} for a review). On the one hand, it is closely related to the one-loop effective action \cite{3}, and on the other hand, it has been proposed to be a candidate for the dark energy \cite{4,5,6}.
Since the advent of string theory, studying physics in higher dimensional spacetime has become a norm rather than an exception.
There have been quite a number of works that explored the Casimir energies of rectangular cavities, spheres and cylinders in higher dimensional spacetimes \cite{7,8,9,10,11,12,13,14,15,16}.

Casimir effect is more interesting when there exist two interacting objects. In the last century, theoretical computations of Casimir interaction were limited to the configuration of two parallel plates. However,  the advancement in nanotechnology and Casimir experiments at the end of the last century have called for theoretical understanding of the Casimir interaction between any non-flat objects. About ten years ago, a major breakthrough in Casimir research was brought by a few groups of researchers \cite{17,18,19,20,21,22,23,24,25,26,27,28,29,30,31,32,33,34,35,36,37,38,39,40}, which have shed new light on the research of Casimir interaction between two objects. Using worldline numerics, multiple-scattering method or mode summation method, exact formulas for the Casimir interactions of cylinder-plate, cylinder-cylinder, sphere-plate, sphere-sphere configurations have been computed. This has enabled the more precise analytical and numerical studies of the nature and the strength of Casimir force. Nonetheless, these works have been limited to the (3+1)-dimensional Minkowski spacetime.

Last year, we have taken the first step to understand the Casimir interactions between non-flat objects in higher dimensional spacetime \cite{41,42,43,44}. We have considered the sphere-plate and sphere-sphere interactions due to the vacuum fluctuations of massless scalar fields in $(D+1)$-dimensional Minkowski spacetime and studied the dependence of the Casimir interaction on the dimension of spacetime. Since cylindrical objects   played an equally important role in physics as spherical objects, we explore the Casimir interaction between a cylinder and a plate, and between two cylinders in $(D+1)$-dimensional Minkowski spacetime in this work.

We generalize the formalism established in \cite{40} to compute the Casimir interaction energy between a cylinder and a plate,  between two parallel cylinders where one lies inside the other, and between two parallel cylinders exterior to each other. We consider massless scalar field with combinations of Dirichlet (D) and Neumann (N) boundary conditions. The generic formula for the Casimir interaction energy between two objects can be written in the form
\begin{equation}
E_{\text{Cas}}=\frac{\hbar}{2\pi}\int_0^{\infty} d\xi\ln\text{Tr}\left(1-\mathbb{T}^1\mathbb{G}^{12}\mathbb{T}^2\mathbb{G}^{21}\right),
\end{equation}   and is thus known as the TGTG formula. Here $\mathbb{T}^i$ is related to the scattering matrix of object $i$ and can be computed by matching the boundary conditions on the object. The matrix $\mathbb{G}^{ij}$  is the translation matrix that relates the wave functions of object $i$ to the wave functions of object $j$. The nontrivial problem is to compute these $\mathbb{T}^i$ and $\mathbb{G}^{ij}$ matrices.

After deriving the TGTG formulas for the cylinder-plate and the cylinder-cylinder interactions, we derive the large separation and small separation asymptotic behaviors of the Casimir interactions. The large separation asymptotic behavior is easy to compute since it only depends on a few entries in each of the $\mathbb{T}$ and $\mathbb{G}$ matrices. To compute the small separation asymptotic behavior beyond the leading term is a tedious task \cite{31,32,41,42,43,44,45,46,47,48,49,50,51}. As a confirmation of the correctness of the TGTG formula, the leading term of the small separation expansion of the Casimir interaction energy is found to agree with that derive using proximity force approximation. One of the major contributions of the present work is the result of the next-to-leading order term of the small separation expansion. For the cylinder-plate configuration with DD boundary conditions, we find that our result agrees with that computed using derivative expansion in \cite{52}. Inspired by the work \cite{53}, we use our results on the cylinder-cylinder interaction to postulate a derivative expansion formula for the Casimir interaction energy in $(D+1)$-dimensional spacetime, up to the second order term, for the interaction between any two objects with combinations of Dirichlet and Neumann boundary conditions. This ansatz is found to agree with the results we derive for the sphere-sphere interaction in \cite{42} except when $D=4$.

This work will  be interesting to those that wish to understand quantum field theory in higher dimensional spacetime.

\section{The Casimir interaction energy}\label{sec2}
In this work, we consider the vacuum fluctuations of a massless scalar field in $(D+1)$-dimensional Minkowski spacetime with metric
\begin{align*}
ds^2=dt^2-dx_1^2-\ldots-dx_D^2.
\end{align*}The equation of motion of the scalar field $\varphi(\mathbf{x})e^{-i\omega t}$, $\mathbf{x}=(x_1,\ldots, x_D)$ is
\begin{equation}\label{eq5_13_1}
\left(\frac{\pa^2}{\pa x_1^2}+\ldots+\frac{\pa^2}{\pa x_D^2}\right)\varphi =-\frac{\omega^2}{c^2}\varphi.
\end{equation}

We will consider the following three problems:
\begin{enumerate}
\item[$\bullet$\;\;] The Casimir interaction between a cylinder and a plate.

\item[$\bullet$\;\;] The Casimir interaction between two parallel cylinders, one is inside the other.

\item[$\bullet$\;\;] The Casimir interaction between two parallel cylinders exterior to each other.

\end{enumerate}The boundary conditions on the cylinder and the plate are either the Dirichlet boundary condition $\varphi\bigr|_{\text{boundary}}=0$ or the Neumann boundary condition $\pa_{\textbf{n}}\varphi\bigr|_{\text{boundary}}=0$, where $\mathbf{n}$ is the unit vector normal to the boundary.

We will take a cylinder to be $$x_1^2+x_2^2=R^2,$$ where $R$ is the radius of the cylinder.
Therefore, it will be convenient to work with the cylindrical coordinates where
\begin{equation}\label{eq5_13_2}x_1=r\cos\theta,\quad x_2=r\sin\theta,\end{equation} so that the cylinder $x_1^2+x_2^2=R^2$ is given by $r=R$.
In the cylindrical coordinates, the equation of motion \eqref{eq5_13_1} reads as
\begin{equation}\label{eq5_13_3}
\left(\frac{\pa^2}{\pa r^2}+\frac{1}{r}\frac{\pa}{\pa r}+\frac{1}{r^2}\frac{\pa^2}{\pa\theta^2}+\frac{\pa^2}{\pa x_3^2}+\ldots+\frac{\pa^2}{\pa x_D^2}\right)\varphi =-\frac{\omega^2}{c^2}\varphi.
\end{equation}
Solving this equation of motion, we find that the cylindrical waves can be parametrized by $(n, k_3,\ldots, k_D)=(n, \boldsymbol{k}_{\perp})$, where $n$ is an integer, $\boldsymbol{k}_{\perp}=(k_3,\ldots,k_D) \in \mathbb{R}^{D-2}$, and they can also be divided into regular and outgoing waves. The explicit formulas for these cylindrical waves are given by
\begin{equation}\label{eq5_13_4}
\begin{split}
\varphi_{n,\boldsymbol{k}_{\perp}}^{*}(\mathbf{x})=\mathcal{C}_n^*Z_n^*(\lambda r) e^{in\theta+ik_3x_3+\ldots +ik_Dx_D},
\end{split}
\end{equation}where $*$ = reg or out for regular or outgoing waves,
\begin{align*}
\mathcal{C}_n^{\text{reg}}=i^{-n},\hspace{1cm}\mathcal{C}_n^{\text{out}}=\frac{\pi}{2}i^{n+1},
\end{align*}are normalization constants, and
\begin{align*}
Z_n^{\text{reg}}(z)=J_n(z),\hspace{1cm}Z_n^{\text{out}}(z)=H_n^{(1)}(z).
\end{align*}

\subsection{The Casimir interaction energy between a cylinder and a plane}
 For the Casimir interaction between a cylinder and a plate, we will take the cylinder to be  $$x_1^2+x_2^2=R^2,\hspace{1cm}-\frac{H_i}{2}\leq x_i\leq \frac{H_i}{2}\quad
 \text{for}\quad 3\leq i\leq D,$$ and the plate to be $$x_1=L,\quad -\frac{H_i}{2}\leq x_i\leq \frac{H_i}{2}\quad \text{for}\quad 2\leq i\leq D.$$
 Here $L>R$ and $d=L-R$ is the distance between the cylinder and the plate.

For the plane $x_1=L$, we will parametrize the plane waves by the momenta perpendicular to the plane $(k_2, k_3,\ldots, k_D)=(k_2,\boldsymbol{k}_{\perp})$. Solving the equation of motion \eqref{eq5_13_1} give the plane wave basis
\begin{equation}\label{eq5_13_7}
\varphi_{k_2,\boldsymbol{k}_{\perp}}^*(\mathbf{x})=e^{-i\text{sgn}_*k_1x_1+ik_2x_2+ik_3x_3+\ldots +ik_Dx_D},
\end{equation}
where
$$k_1=\sqrt{\frac{\omega^2}{c^2}-k_2^2-\ldots-k_D^2},$$   and
\begin{align*}
\text{sgn}_{\text{reg}}=1,\hspace{1cm}\text{sgn}_{\text{out}}=-1.
\end{align*}

Let $H=H_3\ldots H_D$. In the region between the cylinder and the plate, we can write the scalar field $\varphi(\mathbf{x},t)$ in terms of the cylindrical coordinate system centered at the origin:
\begin{equation}\label{eq5_13_5}
\varphi(\mathbf{x}, t)=H\int_{-\infty}^{\infty} d\omega \int_{-\infty}^{\infty}\frac{dk_3}{2\pi}\ldots \int_{-\infty}^{\infty}\frac{dk_D}{2\pi}
\sum_{n=-\infty}^{\infty}\left(a^{n,\boldsymbol{k}_{\perp}}\varphi_{n,\boldsymbol{k}_{\perp}}^{\text{reg}}(\mathbf{x})+b^{n,\boldsymbol{k}_{\perp}}\varphi_{n,\boldsymbol{k}_{\perp}}^{\text{out}}(\mathbf{x})\right)e^{-i\omega t},
\end{equation}or in terms of the rectangular coordinate system centered at $\mathbf{O}'=L\mathbf{e}_1$:
\begin{equation}\label{eq5_13_6}
\varphi(\mathbf{x}', t)=H_2H\int_{-\infty}^{\infty} d\omega \int_{-\infty}^{\infty}\frac{dk_2}{2\pi}\int_{-\infty}^{\infty}\frac{dk_3}{2\pi}\ldots \int_{-\infty}^{\infty}\frac{dk_D}{2\pi}
 \left(c^{k_2,\boldsymbol{k}}\varphi_{k_2,\boldsymbol{k}}^{\text{reg}}(\mathbf{x}')+d^{k_2,\boldsymbol{k}}\varphi_{k_2,\boldsymbol{k}}^{\text{out}}(\mathbf{x}')\right)e^{-i\omega t}.
\end{equation}Here $\mathbf{x}'=\mathbf{x}-L\mathbf{e}_1$.

Using the representation \eqref{eq5_13_5}, we find that the boundary condition on the cylinder $r=R$ gives rise to a relation of the form
\begin{equation}\label{eq5_13_10}
b^{n,\boldsymbol{k}_{\perp}}=-T_1^{n }a^{n,\boldsymbol{k}_{\perp}}.
\end{equation}
For Dirichlet (D) and Neumann (N) boundary conditions, $T_1^{n }$ is given by
\begin{equation}
\begin{split}
T_1^{n, \text{D}}(i\xi)=&\frac{I_n(\gamma R)}{K_n(\gamma R)},\hspace{1cm}
T_1^{n, \text{N}}(i\xi)=\frac{I_n'(\gamma R)}{K_n'(\gamma R)}\end{split}
\end{equation}respectively.
Here $\xi$ is the imaginary frequency so that $i\xi=\omega$,
  $$\gamma=\sqrt{\kappa^2+k_3^2+\ldots+k_D^2},\hspace{1cm}\kappa=\frac{\xi}{c}.$$

Under the representation \eqref{eq5_13_6},   the boundary condition on the plate $x_1'=0$ gives rise to a relation of the form
\begin{equation}\label{eq5_13_11}
c^{k_2,\boldsymbol{k}_{\perp}}=-\widetilde{T}_2^{k_2 }d^{k_2,\boldsymbol{k}_{\perp}}.
\end{equation}
For Dirichlet (D) and Neumann (N) boundary conditions, $\widetilde{T}_2^{k_2 }$ is given by
\begin{equation}
\begin{split}
\widetilde{T}_2^{k_2, \text{D}}(i\xi)=&1,\hspace{1cm}
\widetilde{T}_2^{k_2, \text{D}}(i\xi)=-1\end{split}
\end{equation}respectively.

   The two representations \eqref{eq5_13_5} and \eqref{eq5_13_6} are related by translation matrices $\mathbb{V}$ and $\mathbb{W}$:
\begin{equation}\label{eq5_13_8}\begin{split}
\varphi_{k_2,\boldsymbol{k}_{\perp}}^{\text{reg}}(\mathbf{x}')=&\sum_{n=-\infty}^{\infty} V_{n,k_2}\varphi_{n,\boldsymbol{k}_{\perp}}^{\text{reg}}(\mathbf{x}),\\
\varphi_{n,\boldsymbol{k}_{\perp}}^{\text{out}}(\mathbf{x})=&H\int_{-\infty}^{\infty}\frac{dk_2}{2\pi} W_{k_2,n}\varphi_{k_2,\boldsymbol{k}_{\perp}}^{\text{out}}(\mathbf{x}').
\end{split}\end{equation}It is easy to see that the matrices $\mathbb{V}$ and $\mathbb{W}$ are diagonal in $\mathbf{k}_{\perp}$.
 In fact, canceling out $e^{ik_3x_3+\ldots+ik_Dx_D}$ from both sides, we obtain exactly the same equation as in $D=3$ dimension. Hence, quoting the result from dimension $D=3$ (see for example \cite{40}), we have
 \begin{equation*}
\begin{split}
V_{n,k_2}=&\left(\frac{\sqrt{\gamma^2+k_2^2}+k_2}{\gamma}\right)^ne^{-\sqrt{\gamma^2+k_2^2}L},\\
W_{k_2,n}=&\frac{\pi}{H\sqrt{\gamma^2+k_2^2}}\left(\frac{\sqrt{\gamma^2+k_2^2}+k_2}{\gamma}\right)^ne^{-\sqrt{\gamma^2+k_2^2}L}.
\end{split}
 \end{equation*}
 Notice that
 \eqref{eq5_13_8} implies that
\begin{equation}\label{eq5_13_9}\begin{split}
a^{n,\boldsymbol{k}_{\perp}}=&H_2\int_{-\infty}^{\infty}\frac{dk_2}{2\pi} V_{n,k_2}c^{k_2,\boldsymbol{k}_{\perp}},\\
d^{k_2,\boldsymbol{k}_{\perp}}=&\sum_{n=-\infty}^{\infty} W_{k_2,n}b^{n,\boldsymbol{k}_{\perp}}.\end{split}
\end{equation}

 From \eqref{eq5_13_10}, \eqref{eq5_13_11} and \eqref{eq5_13_9}, we find that
 \begin{equation*}
\begin{split}
b^{n,\boldsymbol{k}_{\perp}}=&-T_1^{n }a^{n,\boldsymbol{k}_{\perp}}\\
=&-T_1^{n }H_2\int_{-\infty}^{\infty}\frac{dk_2}{2\pi} V_{n,k_2}c^{k_2,\boldsymbol{k}_{\perp}}\\
=&T_1^{n }H_2\int_{-\infty}^{\infty}\frac{dk_2}{2\pi} V_{n,k_2}\widetilde{T}_2^{k_2 }d^{k_2,\boldsymbol{k}_{\perp}}\\
=&T_1^{n }H_2\int_{-\infty}^{\infty}\frac{dk_2}{2\pi} V_{n,k_2}\widetilde{T}_2^{k_2 }\sum_{n'=-\infty}^{\infty} W_{k_2,n'}b^{n',\boldsymbol{k}_{\perp}}.
\end{split}
 \end{equation*}This is a relation of the form
 \begin{equation}\label{eq5_13_12}\left(\mathbb{I}-\mathbb{M}\right)\mathbb{B} =0,\end{equation}
 where $$\mathbb{M}=\mathbb{T}_1\mathbb{V}\widetilde{\mathbb{T}}_2\mathbb{W},$$and $\mathbb{B}$ is the column matrix with components $b^{n,\boldsymbol{k}_{\perp}}$.

 The matrix $\mathbb{B}$ must be a nontrivial solution of \eqref{eq5_13_12}. Hence, we obtain the dispersion relation
 $$\det\left(\mathbb{I}-\mathbb{M}\right)=0.$$Using standard contour integration technique, and the fact that all the matrices
 $\mathbb{T}_1$, $\mathbb{V}$, $\widetilde{\mathbb{T}}_2$ and $\mathbb{W}$ are diagonal in $\boldsymbol{k}_{\perp}$, we then find that the Casimir interaction energy is given by
\begin{equation*}\begin{split}
E_{\text{Cas}}=&\frac{\hbar c H}{2\pi}\int_0^{\infty} d\kappa\int_{-\infty}^{\infty}\frac{dk_3}{2\pi}\ldots \int_{-\infty}^{\infty}\frac{dk_D}{2\pi}
\text{Tr}\ln\left(1-\mathbb{M}\right),
\end{split}\end{equation*}
where
\begin{equation*}\begin{split}
M_{n,n'}=&T_1^{n}H_2\int_{-\infty}^{\infty} \frac{dk_2}{2\pi} V_{n,k_2}\widetilde{T}_2^{k_2}W_{k_2,n'}
\\=&T_1^{n}\widetilde{T}_2^{k_2 } \int_{-\infty}^{\infty}\frac{dk_2}{2\sqrt{\gamma^2+k_2^2}}
 \left(\frac{\sqrt{\gamma^2+k_2^2}+k_2}{\gamma}\right)^{n+n'}e^{-2\sqrt{\gamma^2+k_2^2}L}\\
=&T_1^{n}\widetilde{T}_2^{k_2 }K_{n+n'}(2\gamma L).
\end{split}\end{equation*}
Recall that
$$\gamma=\sqrt{\kappa^2+k_{\perp}^2},\quad k_{\perp}=\sqrt{k_3^2+\ldots+k_D^2}.$$
Since the dependence of $M_{n,n'}$ on $(k_3,\ldots, k_D)$ is only through $k_{\perp}$, we have
\begin{equation}\begin{split}
E_{\text{Cas}}=&\frac{\hbar c H}{2\pi}\frac{2\pi^{\frac{D-2}{2}}}{(2\pi)^{D-2}\Gamma\left(\frac{D-2}{2}\right)}
\int_0^{\infty} d\kappa\int_{0}^{\infty}dk_{\perp}\; k_{\perp}^{D-3}
\text{Tr}\ln\left(1-\mathbb{M}\right)\\
=&\frac{\hbar c H}{2^{D-2}\pi^{\frac{D}{2}}\Gamma\left(\frac{D-2}{2}\right)}\int_{0}^{\infty}dk_{\perp}\; k_{\perp}^{D-3}\int_k^{\infty} d\gamma\frac{\gamma}{\sqrt{\gamma^2-k_{\perp}^2}}
\text{Tr}\ln\left(1-\mathbb{M}\right)\\
=&\frac{\hbar c H}{2^{D-2}\pi^{\frac{D}{2}}\Gamma\left(\frac{D-2}{2}\right)}\int_{0}^{\infty}d\gamma\gamma \int_0^{\gamma} dk_{\perp}\frac{k_{\perp}^{D-3}}{\sqrt{\gamma^2-k_{\perp}^2}}
\text{Tr}\ln\left(1-\mathbb{M}\right)\\
=&\frac{\hbar c H}{2^{D-1}\pi^{\frac{D-1}{2}}\Gamma\left(\frac{D-1}{2}\right)}\int_{0}^{\infty}d\gamma\gamma^{D-2}
\text{Tr}\ln\left(1-\mathbb{M}\right).
\end{split}\end{equation}

\subsection{The Casimir interaction energy of one cylinder inside the other}
For the Casimir interaction between two cylinders, one inside the other, we take the smaller cylinder to be $$x_1^2+x_2^2=R_1^2,\hspace{1cm}-\frac{H_i}{2}\leq x_i\leq \frac{H_i}{2}\quad
 \text{for}\quad 3\leq i\leq D,$$ whose center is at the origin $\mathbf{O}$, and the larger cylinder is taken to be
 $$(x_1-L)^2+x_2^2=R_2^2,\hspace{1cm}-\frac{H_i}{2}\leq x_i\leq \frac{H_i}{2}\quad
 \text{for}\quad 3\leq i\leq D,$$ whose center is at $\mathbf{O}'=L\mathbf{e}_1$. Notice that $L<R_2-R_1$ and $d=R_2-R_1-L$ is the distance between the two cylinders.

 As in the cylinder-plate case, we can represent the scalar field $\varphi(\mathbf{x},t)$ in the region between the two cylinders in two different ways, one with respect to the cylindrical coordinate system centered at $\mathbf{O}$:
 \begin{equation}\label{eq5_13_14}
\varphi(\mathbf{x}, t)=H\int_{-\infty}^{\infty} d\omega \int_{-\infty}^{\infty}\frac{dk_3}{2\pi}\ldots \int_{-\infty}^{\infty}\frac{dk_D}{2\pi}
\sum_{n=-\infty}^{\infty}\left(a^{n,\boldsymbol{k}_{\perp}}\varphi_{n,\boldsymbol{k}_{\perp}}^{\text{reg}}(\mathbf{x})+b^{n,\boldsymbol{k}_{\perp}}\varphi_{n,\boldsymbol{k}_{\perp}}^{\text{out}}(\mathbf{x})\right)e^{-i\omega t},
\end{equation}and one is in terms of the cylindrical coordinate system centered at $\mathbf{O}'=L\mathbf{e}_1$:
\begin{equation}\label{eq5_13_15}
\varphi(\mathbf{x}', t)=H\int_{-\infty}^{\infty} d\omega \int_{-\infty}^{\infty}\frac{dk_3}{2\pi}\ldots \int_{-\infty}^{\infty}\frac{dk_D}{2\pi}
\sum_{n=-\infty}^{\infty}\left(c^{n,\boldsymbol{k}_{\perp}}\varphi_{n,\boldsymbol{k}_{\perp}}^{\text{reg}}(\mathbf{x}')+d^{n,\boldsymbol{k}_{\perp}}\varphi_{n,\boldsymbol{k}_{\perp}}^{\text{out}}(\mathbf{x}')\right)e^{-i\omega t}.
\end{equation}

The boundary conditions on the cylinders give
\begin{equation}\label{eq5_13_16}
b^{n,\boldsymbol{k}_{\perp}}=-T_1^{n }a^{n,\boldsymbol{k}_{\perp}},\hspace{1cm} c^{n,\boldsymbol{k}_{\perp}}=-\widetilde{T}_2^{n }d^{n,\boldsymbol{k}_{\perp}},
\end{equation}with
\begin{equation}
\begin{split}
T_1^{n, \text{D}}(i\xi)=&\frac{I_n(\gamma R_1)}{K_n(\gamma R_1)},\hspace{1cm}
T_1^{n, \text{N}}(i\xi)=\frac{I_n'(\gamma R_1)}{K_n'(\gamma R_1)},\\
\widetilde{T}_2^{n, \text{D}}(i\xi)=&\frac{K_n(\gamma R_2)}{I_n(\gamma R_2)},\hspace{1cm}
\widetilde{T}_2^{n, \text{N}}(i\xi)=\frac{K_n'(\gamma R_2)}{I_n'(\gamma R_2)}.\end{split}
\end{equation}
The two representations \eqref{eq5_13_14} and \eqref{eq5_13_15} are related by
\begin{equation}\label{eq5_13_17}\begin{split}
\varphi_{n',\boldsymbol{k}_{\perp}}^{\text{reg}}(\mathbf{x}')=&\sum_{n=-\infty}^{\infty} V_{n,n'}\varphi_{n,\boldsymbol{k}_{\perp}}^{\text{reg}}(\mathbf{x}),\\
\varphi_{n,\boldsymbol{k}_{\perp}}^{\text{out}}(\mathbf{x})=&\sum_{n'=-\infty}W_{n',n}\varphi_{n',\boldsymbol{k}_{\perp}}^{\text{out}}(\mathbf{x}').
\end{split}\end{equation}Compare to the $D=3$ case (see e.g. \cite{40}), we find that
\begin{equation}
\begin{split}
V_{n,n'}=W_{n',n}=(-1)^{n-n'}I_{n'-n}(\gamma L).
\end{split}
\end{equation}
As in the cylinder-plate case, we then find that the Casimir interaction energy is given by
\begin{equation}\label{eq5_13_18}
E_{\text{Cas}}=\frac{\hbar c H}{2^{D-1}\pi^{\frac{D-1}{2}}\Gamma\left(\frac{D-1}{2}\right)}\int_{0}^{\infty}d\gamma\gamma^{D-2}
\text{Tr}\ln\left(1-\mathbb{M}\right),
\end{equation}
where
\begin{equation}\begin{split}
M_{n,n'}=&T_1^{n}\sum_{n''=-\infty}^{\infty}V_{n,n''}\widetilde{T}_2^{n''}W_{n'',n'}\\
=&T_1^{n}\sum_{n''=-\infty}^{\infty}I_{n''-n}T_2^{n''}I_{n''-n'}.
\end{split}\end{equation}

\subsection{The Casimir interaction energy of two parallel cylinder outside each other}
For the Casimir interaction between two parallel cylinders  exterior to each other, we take one cylinder to be $$x_1^2+x_2^2=R_1^2,\hspace{1cm}-\frac{H_i}{2}\leq x_i\leq \frac{H_i}{2}\quad
 \text{for}\quad 3\leq i\leq D,$$ whose center is at the origin $\mathbf{O}$, and the second cylinder is taken to be
 $$(x_1-L)^2+x_2^2=R_2^2,\hspace{1cm}-\frac{H_i}{2}\leq x_i\leq \frac{H_i}{2}\quad
 \text{for}\quad 3\leq i\leq D,$$ whose center is at $\mathbf{O}'=L\mathbf{e}_1$. In this case, $L>R_1+R_2$ and $d=L-R_1-R_2$ is the distance between the two cylinders.

 In the region between the two cylinders, the scalar field $\varphi(\mathbf{x},t)$ can be represented by \eqref{eq5_13_14} using the cylindrical  coordinate system centered at $\mathbf{O}$, or by \eqref{eq5_13_15} using the cylindrical  coordinate system centered at $\mathbf{O}'$.

The boundary conditions on the cylinders give
\begin{equation}\label{eq5_13_20}
b^{n,\boldsymbol{k}_{\perp}}=-T_1^{n }a^{n,\boldsymbol{k}_{\perp}},\hspace{1cm} d^{n,\boldsymbol{k}_{\perp}}=-T_2^{n }c^{n,\boldsymbol{k}_{\perp}},
\end{equation}with
\begin{equation}
\begin{split}
T_i^{n, \text{D}}(i\xi)=&\frac{I_n(\gamma R_i)}{K_n(\gamma R_i)},\hspace{1cm}
T_i^{n, \text{N}}(i\xi)=\frac{I_n'(\gamma R_i)}{K_n'(\gamma R_i)}.\end{split}
\end{equation}In the present case, the two representations \eqref{eq5_13_14} and \eqref{eq5_13_15} are related by
\begin{equation}\label{eq5_13_21}\begin{split}
\varphi_{n',\boldsymbol{k}_{\perp}}^{\text{out}}(\mathbf{x}')=&\sum_{n=-\infty}^{\infty} U^{12}_{n,n'}\varphi_{n,\boldsymbol{k}_{\perp}}^{\text{reg}}(\mathbf{x}),\\
\varphi_{n,\boldsymbol{k}_{\perp}}^{\text{out}}(\mathbf{x})=&\sum_{n'=-\infty}U^{21}_{n',n}\varphi_{n',\boldsymbol{k}_{\perp}}^{\text{reg}}(\mathbf{x}').
\end{split}\end{equation}
In other words,
\begin{equation}\label{eq5_13_22}
\begin{split}
a^{n,\boldsymbol{k}_{\perp}}=&\sum_{n'=-\infty}^{\infty} U_{n,n'}^{12} d^{n',\boldsymbol{k}_{\perp}},\\
c^{n',\boldsymbol{k}_{\perp}}=&\sum_{n=-\infty}^{\infty} U_{n',n}^{21} b^{n,\boldsymbol{k}_{\perp}}.
\end{split}
\end{equation}
Compare to the $D=3$ case (see e.g. \cite{40}), we find that
\begin{equation}
\begin{split}
U_{n,n'}^{12}=U_{n',n}^{21}=(-1)^{n'}K_{n-n'}(\gamma L).
\end{split}
\end{equation}
From \eqref{eq5_13_20} and \eqref{eq5_13_22}, we find that
\begin{equation*}
\begin{split}
b^{n,\boldsymbol{k}_{\perp}}=&-T_1^{n }a^{n,\boldsymbol{k}_{\perp}}\\
=&-T_1^{n }\sum_{n''=-\infty}^{\infty} U_{n,n''}^{12} d^{n'',\boldsymbol{k}_{\perp}}\\
=&T_1^{n }\sum_{n''=-\infty}^{\infty} U_{n,n''}^{12}  T_2^{n''}c^{n'',\boldsymbol{k}_{\perp}}\\
=&T_1^{n }\sum_{n''=-\infty}^{\infty} U_{n,n''}^{12}  T_2^{n''}\sum_{n'=-\infty}^{\infty} U_{n'',n'}^{21} b^{n',\boldsymbol{k}_{\perp}}.
\end{split}
 \end{equation*}

Then as in the cylinder-plate case, we find that the Casimir interaction energy is given by
\begin{equation}\label{eq5_13_23}
E_{\text{Cas}}=\frac{\hbar c H}{2^{D-1}\pi^{\frac{D-1}{2}}\Gamma\left(\frac{D-1}{2}\right)}\int_{0}^{\infty}d\gamma\gamma^{D-2}
\text{Tr}\ln\left(1-\mathbb{M}\right),
\end{equation}
where
\begin{equation}\label{eq5_13_25}\begin{split}
M_{n,n'}=&T_1^{n}\sum_{n''=-\infty}^{\infty}U_{n,n''}^{12}T_2^{n''}U_{n'',n'}^{21}\\
=&T_1^{n}\sum_{n''=-\infty}^{\infty}K_{n-n''}T_2^{n''}K_{n'-n''}.
\end{split}\end{equation}
Using the fact that $I_{-n}(z)=I_n(z)$, $K_{-n}(z)=K_n(z)$, we find that $T_2^{-n}=T_2^{n}$. Hence, \eqref{eq5_13_25} can be rewritten as
\begin{equation}\label{eq5_13_26}\begin{split}
M_{n,n'}=&T_1^{n}\sum_{n''=-\infty}^{\infty}U_{n,n''}^{12}T_2^{n''}U_{n'',n'}^{21}\\
=&T_1^{n}\sum_{n''=-\infty}^{\infty}K_{n+n''}T_2^{n''}K_{n'+n''}.
\end{split}\end{equation}

\section{ Large separation asymptotic behavior  }
In this section, we compute the asymptotic behavior of the Casimir interaction energy when $L\gg R$ and $L\gg R_1+R_2$, for the cylinder-plate interaction and the cylinder-cylinder interaction when the two cylinders are outside each other.

First notice that  by making a change of variables $\gamma=\tilde{\gamma}/L$ and expanding the logarithm, we have
 \begin{equation}\label{eq5_13_24}\begin{split}
E_{\text{Cas}}=&\frac{\hbar c H}{2^{D-1}\pi^{\frac{D-1}{2}}\Gamma\left(\frac{D-1}{2}\right)L^{D-1}}\int_{0}^{\infty}d\tilde{\gamma}\tilde{\gamma}^{D-2}
\text{Tr}\ln\left(1-\mathbb{M}\right)\\
=&-\frac{\hbar c H}{2^{D-1}\pi^{\frac{D-1}{2}}\Gamma\left(\frac{D-1}{2}\right)L^{D-1}}\sum_{s=0}^{\infty}\frac{1}{s+1}\int_{0}^{\infty}d\tilde{\gamma}\tilde{\gamma}^{D-2}
\sum_{n_0=-\infty}^{\infty}\ldots\sum_{n_s=-\infty}^{\infty}M_{n_0,n_1}\ldots M_{n_s, n_0}.
\end{split}\end{equation}
For the cylinder-plate case,
\begin{equation*}
M_{n,n'}^{\text{cp}}=T^n_1\widetilde{T}_2^{k_2}K_{n+n'}(2\tilde{\gamma}).
\end{equation*}For the case of two cylinders exterior to each other,
\begin{equation*}
M_{n,n'}^{\text{cc}}=T^n_1\sum_{n''=-\infty}^{\infty} K_{n+n''}(\tilde{\gamma})T_2^{n''}K_{n'+n''}(\tilde{\gamma})=\sum_{n''=-\infty}^{\infty}N_{n,n''}^1N_{n'',n'}^2.
\end{equation*}Here
\begin{equation*}
T_i^{n, \text{D}}=\frac{I_n\left(\tilde{\gamma}R_i/L\right)}{K_n\left(\tilde{\gamma}R_i/L\right)},\hspace{1cm}T_i^{n, \text{N}}=\frac{I_n'\left(\tilde{\gamma}R_i/L\right)}{K_n'\left(\tilde{\gamma}R_i/L\right)},
\end{equation*}
and
\begin{align*}
N_{n,n''}^i=T^n_iK_{n+n''}(\tilde{\gamma}).
\end{align*}
From these, we see that to determine the large separation asymptotic behavior (i.e. $L\gg R$ in the cylinder-plate case and $L\gg R_1+R_2$ in the cylinder-cylinder case) of the Casimir interaction energy, we need to find the small $z$ asymptotic behavior of $I_n(z)/K_n(z)$ and   $I_n'(z)/K_n'(z)$.

From any standard textbook of special functions, we find that
 \begin{equation}\label{eq5_13_26}\begin{split}
\frac{I_0(z)}{K_0(z)}=&-\frac{1}{\ln z}+\ldots,\\
\frac{I_0'(z)}{K_0'(z)}=&-\frac{1}{2}z^2+\ldots,\\
\frac{I_1'(z)}{K_1'(z)}=&-\frac{1}{2}z^{2}+\ldots,\\
\frac{I_n(z)}{K_n(z)}=&O\left(z^{2n}\right),\quad n\geq 1,\\
\frac{I_n'(z)}{K_n'(z)}=&O\left(z^{2n}\right),\quad n\geq 1.
\end{split}
\end{equation}

Hence, for the cylinder-plate case, we find that when the cylinder is imposed with Dirichlet boundary conditions, the leading term of the large separation asymptotic expansion comes from the term $s=0$ and $n_0=0$; whereas when the cylinder is imposed with the Neumann boundary conditions,  the leading term of the large separation asymptotic expansion comes from the term $s=0$ and $n_0=0,\pm 1$, namely,
\begin{equation*}\begin{split}
 E_{\text{Cas}}^{\text{DD}} \sim&-\frac{\hbar c H}{2^{D-1}\pi^{\frac{D-1}{2}}\Gamma\left(\frac{D-1}{2}\right)L^{D-1}}\int_{0}^{\infty}d\tilde{\gamma}\tilde{\gamma}^{D-2}M_{0,0}^{\text{cp}, \text{DD}},\\
  E_{\text{Cas}}^{\text{DN}} \sim &-\frac{\hbar c H}{2^{D-1}\pi^{\frac{D-1}{2}}\Gamma\left(\frac{D-1}{2}\right)L^{D-1}}\int_{0}^{\infty}d\tilde{\gamma}\tilde{\gamma}^{D-2}M_{0,0}^{\text{cp}, \text{DN}},\\
   E_{\text{Cas}}^{\text{ND}} \sim &-\frac{\hbar c H}{2^{D-1}\pi^{\frac{D-1}{2}}\Gamma\left(\frac{D-1}{2}\right)L^{D-1}}\int_{0}^{\infty}d\tilde{\gamma}\tilde{\gamma}^{D-2}
   \left(M_{0,0}^{\text{cp}, \text{ND}}+M_{1,1}^{\text{cp}, \text{ND}}+M_{-1,-1}^{\text{cp}, \text{ND}}\right),\\
    E_{\text{Cas}}^{\text{NN}} \sim &-\frac{\hbar c H}{2^{D-1}\pi^{\frac{D-1}{2}}\Gamma\left(\frac{D-1}{2}\right)L^{D-1}}\int_{0}^{\infty}d\tilde{\gamma}\tilde{\gamma}^{D-2}
   \left(M_{0,0}^{\text{cp}, \text{NN}}+M_{1,1}^{\text{cp}, \text{NN}}+M_{-1,-1}^{\text{cp}, \text{NN}}\right).
\end{split}\end{equation*}Straightforward computation then gives
\begin{equation}\label{eq5_13_28}\begin{split}
 E_{\text{Cas}}^{\text{DD}} \sim &-\frac{\hbar c H\Gamma\left(\frac{D-1}{2}\right)}{2^{D+1}\pi^{\frac{D-1}{2}}L^{D-1}\ln\left(L/R\right)},
\\
E_{\text{Cas}}^{\text{DN}} \sim &\frac{\hbar c H\Gamma\left(\frac{D-1}{2}\right)}{2^{D+1}\pi^{\frac{D-1}{2}}L^{D-1}\ln\left(L/R\right)},
\\
E_{\text{Cas}}^{\text{ND}} \sim & \frac{\hbar c H(3D+1)\Gamma\left(\frac{D+1}{2}\right) R^2}{2^{D+3}\pi^{\frac{D-1}{2}}L^{D+1}},\\
E_{\text{Cas}}^{\text{NN}} \sim & -\frac{\hbar c H(3D+1)\Gamma\left(\frac{D+1}{2}\right) R^2}{2^{D+3}\pi^{\frac{D-1}{2}}L^{D+1}}.
\end{split}\end{equation}
Notice that if Dirichlet boundary condition is imposed on the cylinder, the leading term is of order $L^{-D+1}/\ln (L)$; whereas if Neumann boundary condition is imposed on the cylinder, the leading term is of order $L^{-D-1}$.

When $D=3$, \eqref{eq5_13_28} reads as
\begin{equation}\label{eq5_13_29}\begin{split}
 E_{\text{Cas}}^{\text{DD}} \sim &-\frac{\hbar c H }{16\pi L^{2}\ln\left(L/R\right)},
\\
E_{\text{Cas}}^{\text{NN}} \sim & -\frac{5\hbar c H  R^2}{32\pi L^4}.
\end{split}\end{equation}These agree with the results obtained in \cite{24}.

For two cylinders that are exterior to each other, we rewrite \eqref{eq5_13_24} as
 \begin{equation}\label{eq5_13_24_2}\begin{split}
E_{\text{Cas}}=& -\frac{\hbar c H}{2^{D-1}\pi^{\frac{D-1}{2}}\Gamma\left(\frac{D-1}{2}\right)L^{D-1}}\\&\hspace{1cm}\times\sum_{s=0}^{\infty}\frac{1}{s+1}\int_{0}^{\infty}d\tilde{\gamma}\tilde{\gamma}^{D-2}
\sum_{n_0=-\infty}^{\infty}\ldots\sum_{n_s=-\infty}^{\infty}\sum_{n_0'=-\infty}^{\infty}\ldots\sum_{n_s'=-\infty}^{\infty}N_{n_0,n_0'}^1N_{n_0',n_1}^2\ldots N_{n_s,n_s'}^1N_{n_s',n_0}^2.
\end{split}\end{equation}
As in the cylinder-plate case, we find that the leading terms of the large separation asymptotic expansions are given by
\begin{equation*}\begin{split}
 E_{\text{Cas}}^{\text{DD}} \sim &-\frac{\hbar c H}{2^{D-1}\pi^{\frac{D-1}{2}}\Gamma\left(\frac{D-1}{2}\right)L^{D-1}}\int_{0}^{\infty}d\tilde{\gamma}\tilde{\gamma}^{D-2}N_{0,0}^{1,\text{D}}N_{0,0}^{2,\text{D}},\\
  E_{\text{Cas}}^{\text{DN}} \sim &-\frac{\hbar c H}{2^{D-1}\pi^{\frac{D-1}{2}}\Gamma\left(\frac{D-1}{2}\right)L^{D-1}}\int_{0}^{\infty}d\tilde{\gamma}\tilde{\gamma}^{D-2}
  \left(N_{0,0}^{1,\text{D}}N_{0,0}^{2,\text{N}}+N_{0,1}^{1,\text{D}}N_{1,0}^{2,\text{N}}+N_{0,-1}^{1,\text{D}}N_{-1,0}^{2,\text{N}}\right),\\
   E_{\text{Cas}}^{\text{NN}} \sim &-\frac{\hbar c H}{2^{D-1}\pi^{\frac{D-1}{2}}\Gamma\left(\frac{D-1}{2}\right)L^{D-1}}\int_{0}^{\infty}d\tilde{\gamma}\tilde{\gamma}^{D-2}
  \left( N_{0,0}^{1,\text{N}}N_{0,0}^{2,\text{N}}+N_{0,1}^{1,\text{N}}N_{1,0}^{2,\text{N}}+N_{0,-1}^{1,\text{N}}N_{-1,0}^{2,\text{N}}
  \right.\\&\left.+N_{-1,0}^{1,\text{N}}N_{0,-1}^{2,\text{N}}+N_{1,0}^{1,\text{N}}N_{0,1}^{2,\text{N}}++N_{1,1}^{1,\text{N}}N_{1,1}^{2,\text{N}}
  +N_{-1,-1}^{1,\text{N}}N_{-1,-1}^{2,\text{N}}+N_{1,-1}^{1,\text{N}}N_{-1,1}^{2,\text{N}}++N_{-1,1}^{1,\text{N}}N_{1,-1}^{2,\text{N}}\right),
\end{split}\end{equation*}and $E_{\text{Cas}}^{\text{ND}} $ is obtained from $E_{\text{Cas}}^{\text{DN}} $ by interchanging $R_1$ and $R_2$.

Straightforward computation gives
\begin{equation}\begin{split}
 E_{\text{Cas}}^{\text{DD}} \sim &-\frac{\hbar c H\Gamma\left(\frac{D-1}{2}\right)^2}{2^{D+1}\pi^{\frac{D-2}{2}}\Gamma\left(\frac{D}{2}\right)L^{D-1}\ln\left(L/R_1\right)
 \ln\left(L/R_2\right)},\\
E_{\text{Cas}}^{\text{DN}} \sim & \frac{\hbar c \Gamma\left(\frac{D+1}{2}\right)^2(3D+1)H R_2^2}{2^{D+3}\pi^{\frac{D-2}{2}}\Gamma\left(\frac{D+2}{2}\right)L^{D+1}\ln\left(L/R_1\right)},
\\
   E_{\text{Cas}}^{\text{NN}} \sim & -\frac{3\hbar c(D+1)(3D+5)\Gamma \left(\frac{D+3}{2}\right)^2 H R_1^2R_2^2}{2^{D+5}\pi^{\frac{D-2}{2}}\Gamma\left(\frac{D+4}{2}\right)L^{D+3}}.
\end{split}\end{equation}Notice that the leading term of the Casimir interaction is of order $L^{-D+1}/(\ln L)^2$, $L^{-D-1}/(\ln L)$ and $L^{-D-3}$ respectively for DD, DN and NN boundary conditions.

From above, we see that at large separation, the decay of the Casimir interaction   is slower when Dirichlet boundary condition is imposed on the cylinder, and the decay is faster when Neumann boundary conditions is imposed on the cylinder. For the same boundary conditions, the decay is faster in higher dimensions.

\section{Small separation asymptotic behavior}
Small separation asymptotic behavior of the Casimir interaction is of much more interest since Casimir force is inversely proportional to some power of the distance between the objects. It is always expected that the leading term of the Casimir interaction should agree with that derived using the proximity force approximation. A subject of much more interest is the next-to-leading order term, because the ratio of the next-to-leading order term to the leading order term is experimentally measurable.   For the cylinder-plate configuration in $(3+1)$ dimensions, the small
separation asymptotic behavior has been derived in \cite{31} up to the next-to-leading order term. The idea of the derivation is to similar to perturbation in quantum field theory. Later the method has been generalized to compute the small separation asymptotic behavior of the Casimir interaction energy in various settings \cite{32, 41, 42, 44,45,46,47,48,49,50,51}.

In Section \ref{sec2}, we have seen that for the cylinder-plate or cylinder-cylinder configurations, the Casimir interaction energy is given by
\begin{equation}\label{eq5_14_1}
E_{\text{Cas}}=\frac{\hbar c H}{2^{D-1}\pi^{\frac{D-1}{2}}\Gamma\left(\frac{D-1}{2}\right)}\int_{0}^{\infty}d\gamma\gamma^{D-2}
\text{Tr}\ln\left(1-\mathbb{M}\right),
\end{equation}with different matrix $\mathbb{M}$ for different scenarios.
The first step in deriving the small separation asymptotic expansion is to expand the logarithm in \eqref{eq5_14_1}, which gives
\begin{equation}\label{eq5_14_2}
E_{\text{Cas}}=-\frac{\hbar c H}{2^{D-1}\pi^{\frac{D-1}{2}}\Gamma\left(\frac{D-1}{2}\right)}\sum_{s=0}^{\infty}\frac{1}{s+1}\int_{0}^{\infty}d\gamma\gamma^{D-2}
\sum_{n_0=-\infty}^{\infty}\ldots\sum_{n_s=-\infty}^{\infty}M_{n_0,n_1}\ldots M_{n_s, n_0}.
\end{equation}
In the following, we will discuss the different scenarios separately.

\subsection{The cylinder-plate case}\label{sec4_1}
\begin{figure}[h]
\epsfxsize=0.4\linewidth \epsffile{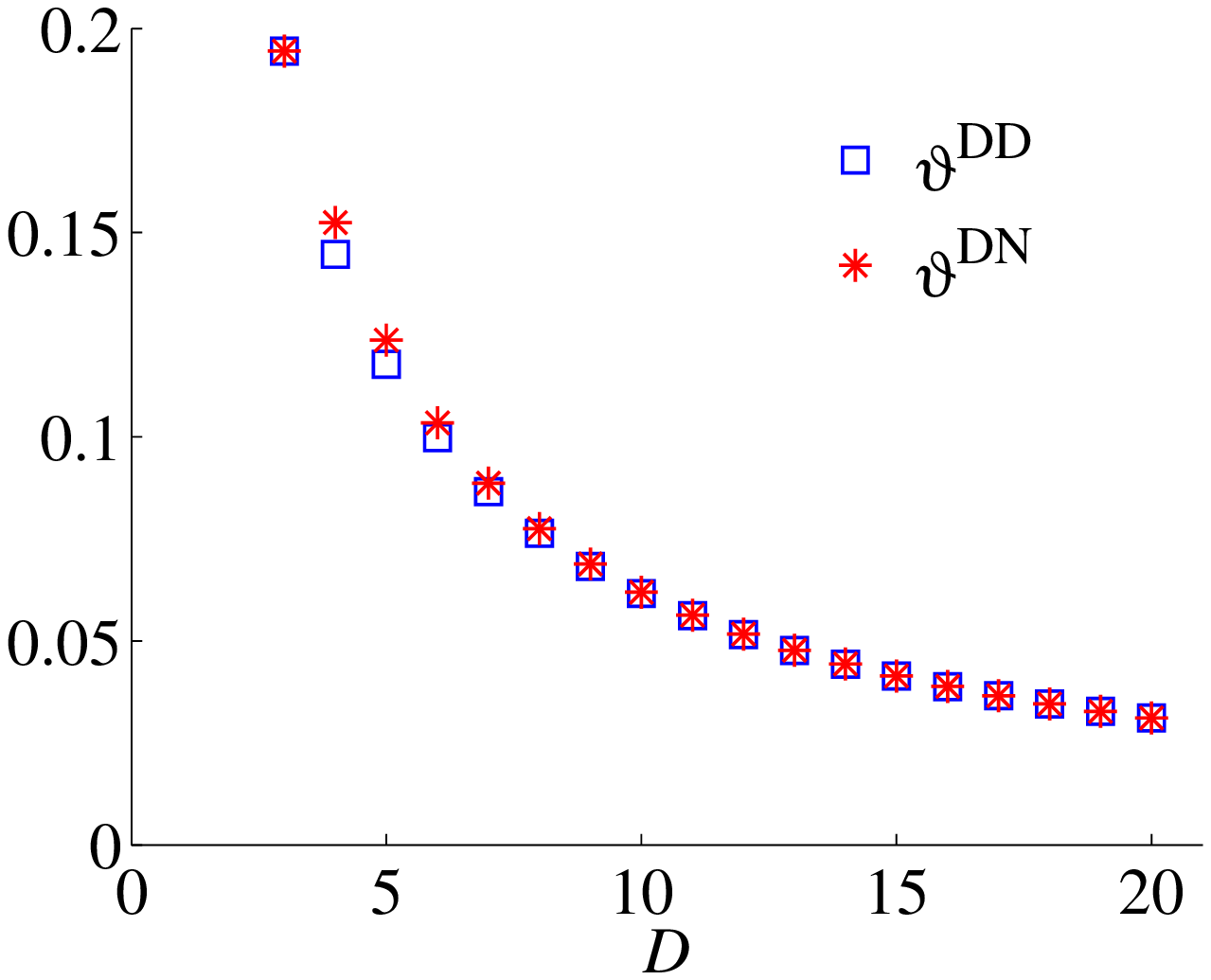} \epsfxsize=0.4\linewidth \epsffile{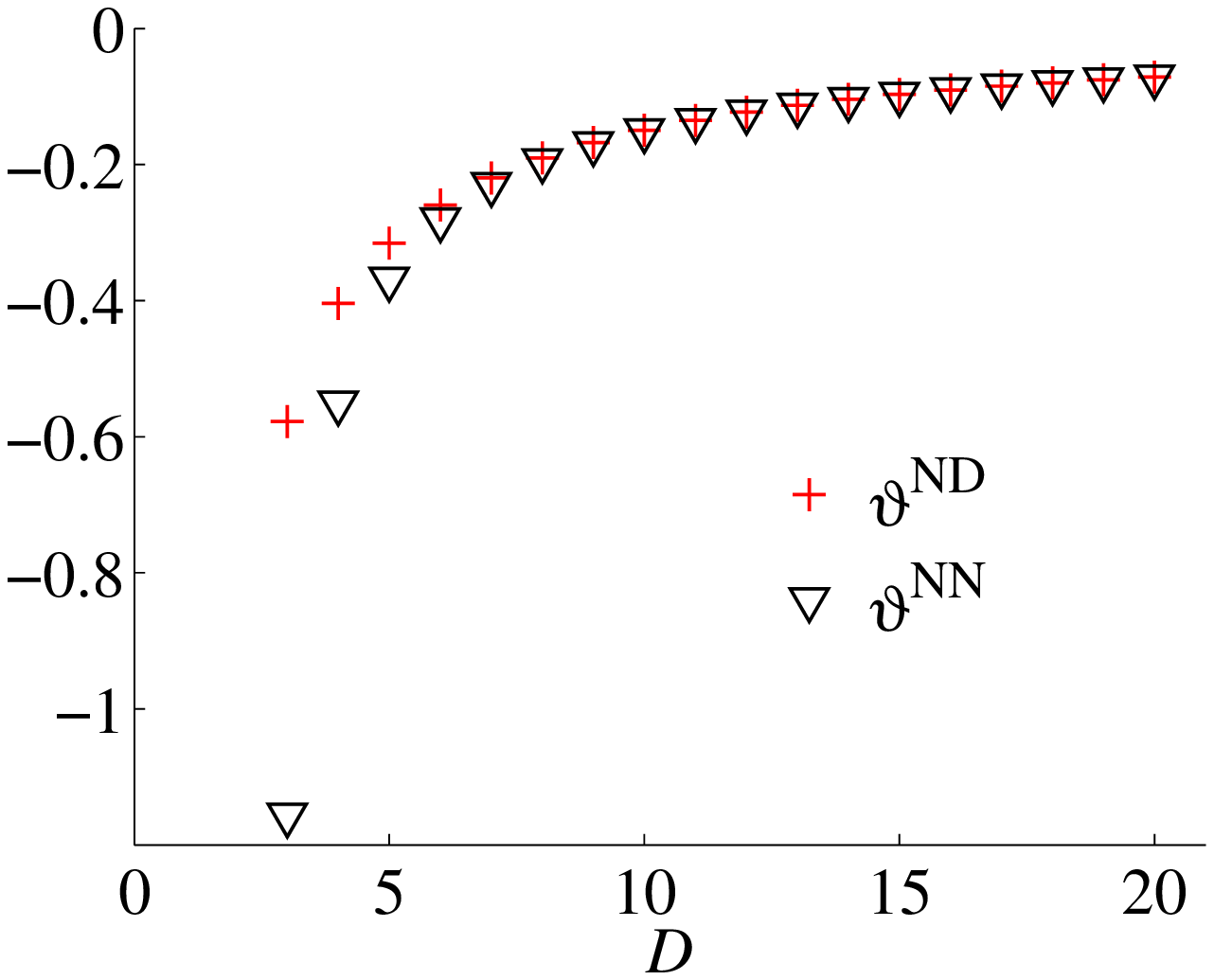}
 \caption{\label{f1} The dependence of $\vartheta$  on dimension $D$ for different combinations of boundary conditions.}\end{figure}

In this case,
define
\begin{equation*}
\vep=\frac{d}{R},\quad n=n_0,\quad \omega=\gamma R,
\end{equation*}and make a change of variables
\begin{gather*}
n_i=n+\tilde{n}_i,\quad 1\leq i\leq s,\\
\omega=\frac{n\sqrt{1-\tau^2}}{\tau}.
\end{gather*}Approximating the summation by corresponding integrations, we have
\begin{equation}\label{eq5_14_2}\begin{split}
E_{\text{Cas}}\sim &-\frac{\hbar c H}{2^{D-2}\pi^{\frac{D-1}{2}}\Gamma\left(\frac{D-1}{2}\right)R^{D-1}}\sum_{s=0}^{\infty}\frac{1}{s+1}  \int_{0}^{\infty}
dn \, n^{D-1}\int_0^1 d\tau\frac{(1-\tau^2)^{\frac{D-3}{2}}}{\tau^{D}}\int_{-\infty}^{\infty} d\tilde{n}_1\ldots\int_{-\infty}^{\infty} d\tilde{n}_s
M_{n_0,n_1}\ldots M_{n_{s},n_0},
\end{split}\end{equation}
where
\begin{equation*}
\begin{split}
M_{n_i,n_{i+1}}^{\text{XY}}=(-1)^{\alpha_{\text{Y}}}T_1^{n_i, \text{X}} K_{n_i+n_{i+1}}\left(2\omega (1+\vep)\right).
\end{split}
\end{equation*}Here X $=$ D or N is the boundary condition on the sphere, and Y $=$ D or N is the boundary condition on the plate,
$\alpha_{\text{D}}=0$, $\alpha_{\text{N}}=1$,
\begin{equation*}
\begin{split}
T_1^{n_i, \text{D}}=\frac{I_{n_i}(\omega)}{K_{n_i}(\omega)},\hspace{1cm}T_1^{n_i, \text{N}}=\frac{I_{n_i}'(\omega)}{K_{n_i}'(\omega)}.
\end{split}
\end{equation*}
Now we need to find the asymptotic expansion of $M_{n_i,n_{i+1}}$ keeping in mind that $n\sim \vep^{-1}$, $\tilde{n}_i\sim \vep^{-\frac{1}{2}}$. We also need the Debye asymptotic expansions of the modified Bessel functions given below:
 \begin{equation}\label{eq3_26_6}\begin{split}
I_{\nu}(\nu z)\sim & \frac{1}{\sqrt{2\pi \nu}}\frac{e^{\nu\eta(z)}}{(1+z^2)^{\frac{1}{4}}}\left(1+\frac{u_1(t(z))}{\nu}+\ldots\right),\\
K_{\nu}(\nu z)\sim &\sqrt{\frac{\pi}{ 2 \nu}}\frac{e^{-\nu\eta(z)}}{(1+z^2)^{\frac{1}{4}}}\left(1-\frac{u_1(t(z))}{\nu}+\ldots\right),\\
I_{\nu}'(\nu z)\sim & \frac{1}{\sqrt{2\pi \nu}}\frac{e^{\nu\eta(z)}(1+z^2)^{\frac{1}{4}}}{z}\left(1+\frac{v_1(t(z))}{\nu}+\ldots\right),\\
K_{\nu}'(\nu z)\sim &-\sqrt{\frac{\pi}{ 2 \nu}}\frac{e^{-\nu\eta(z)}(1+z^2)^{\frac{1}{4}}}{z}\left(1-\frac{v_1(t(z))}{\nu}+\ldots\right),
\end{split}\end{equation}where
\begin{gather}
\eta(z)= \sqrt{1+z^2}+\log\frac{z}{1+\sqrt{1+z^2}},\hspace{1cm}
t(z)=\frac{1}{\sqrt{1+z^2}},\\
u_1(t)=\frac{t}{8}-\frac{5t^3}{24},\hspace{1cm}
v_1(t)=-\frac{3t}{8}+\frac{7t^3}{24}.
\end{gather}
From this, we find that
\begin{equation*}\begin{split}
\frac{I_{\nu}(\nu z)}{K_{\nu}(\nu z)}\sim & \frac{1}{\pi} e^{2\nu\eta(z)}\left(1+\frac{2u_1(t(z))}{\nu}+\ldots\right),\\
\frac{I_{\nu}'(\nu z)}{K_{\nu}'(\nu z)}\sim & -\frac{1}{\pi} e^{2\nu\eta(z)}\left(1+\frac{2v_1(t(z))}{\nu}+\ldots\right).
\end{split}\end{equation*}Therefore,
\begin{equation*}
M_{n_i,n_{i+1}}^{\text{XY}}\sim   (-1)^{\alpha_{\text{X}}+\alpha_{\text{Y}}}\frac{1}{\sqrt{2\pi\nu_2}} \frac{e^{2\nu_1\eta(z_1)-\nu_2\eta(z_2)}}{(1+z_2^2)^{\frac{1}{4}}}\left(1+\mathcal{A}_2^{X}\right),
\end{equation*}where
\begin{gather*}
\nu_1=n_i,\quad \nu_2=n_i+n_{i+1}\\
z_1=\frac{\omega}{n_i},\quad z_2=\frac{2\omega(1+\vep)}{n_i+n_{i+1}},
\end{gather*} and
\begin{equation*}\begin{split}
\mathcal{A}_2^{\text{D}}=&\frac{1}{n}\left(2u_1(\tau)-\frac{1}{2}u_1(\tau)\right),\\
\mathcal{A}_2^{\text{N}}=&\frac{1}{n}\left(2v_1(\tau)-\frac{1}{2}u_1(\tau)\right)
\end{split}\end{equation*}are of order $\vep$.
With the help of a computer symbolic math package, we find that
\begin{equation*}\begin{split}
M_{n_i,n_{i+1}}^{\text{XY}}
\sim &  (-1)^{\alpha_{\text{X}}+\alpha_{\text{Y}}}C^{\tilde{n}_i-\tilde{n}_{i+1}}\frac{1}{2 } \sqrt{\frac{\tau}{\pi n}}\left(1+\mathcal{B}_{i,1}+\mathcal{B}_{i,2}\right)
\exp\left(-\frac{2\vep n}{\tau}-\frac{\tau\left(\tilde{n}_i-\tilde{n}_{i+1}\right)^2}{4n} \right)\left(1+\mathcal{A}_2^{X}\right),
\end{split}\end{equation*}where $\mathcal{B}_{i,1}$ and $\mathcal{B}_{i,2}$ are respectively terms of order $\sqrt{\vep}$ and $\vep$.
Substitute into \eqref{eq5_14_2}, we find that
\begin{equation*}\begin{split}
E_{\text{Cas}}^{\text{XY}}\sim &-\frac{\hbar c H}{2^{D-1}\pi^{\frac{D}{2}}\Gamma\left(\frac{D-1}{2}\right)R^{D-1}}\sum_{s=0}^{\infty}\frac{ (-1)^{(\alpha_{\text{X}}+\alpha_{\text{Y}})(s+1)}}{s+1} \frac{1}{2^s\pi^{\frac{s}{2}}}\int_{0}^{\infty}
dn \, n^{D-1-\frac{s+1}{2}}\int_0^1 d\tau\frac{(1-\tau^2)^{\frac{D-3}{2}}}{\tau^{D-\frac{s+1}{2}}}\int_{-\infty}^{\infty} d\tilde{n}_1\ldots\int_{-\infty}^{\infty} d\tilde{n}_s\\
&\times \exp\left(-\frac{2\vep (s+1) n}{\tau}-\sum_{i=0}^s\frac{\tau\left(\tilde{n}_i-\tilde{n}_{i+1}\right)^2}{4n} \right)\left(1+(s+1)\mathcal{A}_2^{X}\right)
\left(1+\sum_{i=0}^{s}\mathcal{B}_{i,1}+\sum_{i=0}^{s-1}\sum_{j=i+1}^s\mathcal{B}_{i,1}\mathcal{B}_{j,1}+\sum_{i=0}^s\mathcal{B}_{i,2}\right).\\
\end{split}\end{equation*}
The integration over $\tilde{n}_i$ is Gaussian, and it has been explained in \cite{31} (see also \cite{49}). One finds that the terms of order $\sqrt{\vep}$ would not contribute since it is odd in one of the $\tilde{n}_i$. After the integration, one is left with an expression of the form
\begin{equation*}\begin{split}
E_{\text{Cas}}^{\text{XY}}\sim&-\frac{\hbar c H}{2^{D-1}\pi^{\frac{D}{2}}\Gamma\left(\frac{D-1}{2}\right)R^{D-1}}\sum_{s=0}^{\infty}\frac{(-1)^{(\alpha_{\text{X}}+\alpha_{\text{Y}})(s+1)}}{(s+1)^{\frac{3}{2}}}\int_0^1 d\tau\frac{(1-\tau^2)^{\frac{D-3}{2}}}{\tau^{D-\frac{ 1}{2}}}  \int_{0}^{\infty}
dn \, n^{D- \frac{3}{2}}\exp\left(-\frac{2\vep (s+1) n}{\tau} \right)\left(1 +\mathcal{F}^{\text{X}}\right),\end{split}\end{equation*}
where $\mathcal{F}$ is a term of order $\vep$. The integration over $n$ is straightforward using the definition of gamma function. One obtain
\begin{equation*}\begin{split}
E_{\text{Cas}}^{\text{XY}}\sim &-\frac{\hbar c \Gamma\left(D-\frac{1}{2}\right)H\sqrt{R}}{2^{2D-\frac{3}{2}} \pi^{\frac{D}{2}}\Gamma\left(\frac{D-1}{2}\right)d^{D-\frac{1}{2}}}\sum_{s=0}^{\infty}
\frac{(-1)^{(\alpha_{\text{X}}+\alpha_{\text{Y}})(s+1)}}{(s+1)^{D+1}}  \int_0^1 d\tau (1-\tau^2)^{\frac{D-3}{2}} \left(1+ \mathcal{G}^{\text{X}}\right),
\end{split}\end{equation*}
where $\mathcal{G}^{\text{X}}$ is a term of order $\vep$ and is a polynomial of degree two in $\tau^2$. Using
\begin{equation*}\begin{split}
\int_0^1d\tau\,\tau^{\alpha} (1-\tau^2)^{\frac{D-3}{2}}= \frac{1}{2}\frac{\Gamma\left(\frac{D-1}{2}\right)\Gamma\left(\frac{1+\alpha}{2}\right)}
{\Gamma\left(\frac{D+\alpha}{2}\right)},
\end{split}\end{equation*}
we find that
\begin{equation*}\begin{split}
E_{\text{Cas}}^{\text{XY}}\sim &-\frac{\hbar c \Gamma\left(D-\frac{1}{2}\right)H\sqrt{R}}{2^{2D-\frac{1}{2}} \pi^{\frac{D-1}{2}}\Gamma\left(\frac{D}{2}\right)d^{D-\frac{1}{2}}}\sum_{s=0}^{\infty}\frac{(-1)^{(\alpha_{\text{X}}+\alpha_{\text{Y}})(s+1)}}{(s+1)^{D+1}}
\left(1+ \mathcal{H}^{\text{X}}\right),
\end{split}\end{equation*}
where
\begin{equation*}
\begin{split}
\mathcal{H}^{\text{D}}=&\left(\frac{4D-5}{12(2D-3)}-\frac{(D-2)(D-3)}{3D(2D-3)}(s+1)^2\right)\frac{d}{R},\\
\mathcal{H}^{\text{N}}=&\left(\frac{4D-5}{12(2D-3)}-\frac{D^2+7D-6}{3D(2D-3)}(s+1)^2\right)\frac{d}{R}.
\end{split}
\end{equation*}
Finally using the fact that
\begin{equation*}\begin{split}
\sum_{s=0}^{\infty} \frac{1}{(s+1)^k}=\zeta(k),\quad \sum_{s=0}^{\infty} \frac{(-1)^{s+1}}{(s+1)^k}=-\left(1-2^{-k+1}\right)\zeta(k),
\end{split}\end{equation*}we have
\begin{equation}\label{eq5_18_1}\begin{split}
E_{\text{Cas}}^{\text{DD}}\sim & -\frac{\hbar c \Gamma\left(D-\frac{1}{2}\right)\zeta(D+1)H\sqrt{R}}{2^{2D-\frac{1}{2}} \pi^{\frac{D-1}{2}}\Gamma\left(\frac{D}{2}\right)d^{D-\frac{1}{2}}}
\left(1+ \left[\frac{4D-5}{12(2D-3)}-\frac{(D-2)(D-3)}{3D(2D-3)} \frac{\zeta(D-1)}{\zeta(D+1)}\right]\frac{d}{R}+\ldots \right),\\
E_{\text{Cas}}^{\text{DN}}\sim & (1-2^{-D})  \frac{\hbar c \Gamma\left(D-\frac{1}{2}\right)\zeta(D+1)H\sqrt{R}}{2^{2D-\frac{1}{2}} \pi^{\frac{D-1}{2}}\Gamma\left(\frac{D}{2}\right)d^{D-\frac{1}{2}}}
\left(1+ \left[\frac{4D-5}{12(2D-3)}-\frac{(D-2)(D-3)}{3D(2D-3)} \frac{2^D-4}{2^D-1}\frac{\zeta(D-1)}{\zeta(D+1)}\right]\frac{d}{R}+\ldots \right),\\
E_{\text{Cas}}^{\text{ND}}\sim & (1-2^{-D})  \frac{\hbar c \Gamma\left(D-\frac{1}{2}\right)\zeta(D+1)H\sqrt{R}}{2^{2D-\frac{1}{2}} \pi^{\frac{D-1}{2}}\Gamma\left(\frac{D}{2}\right)d^{D-\frac{1}{2}}}
\left(1+ \left[\frac{4D-5}{12(2D-3)}-\frac{D^2+7D-6}{3D(2D-3)}\frac{2^D-4}{2^D-1}\frac{\zeta(D-1)}{\zeta(D+1)}\right]\frac{d}{R}+\ldots \right),\\
E_{\text{Cas}}^{\text{NN}}\sim & - \frac{\hbar c \Gamma\left(D-\frac{1}{2}\right)\zeta(D+1)H\sqrt{R}}{2^{2D-\frac{1}{2}} \pi^{\frac{D-1}{2}}\Gamma\left(\frac{D}{2}\right)d^{D-\frac{1}{2}}} \zeta(D+1)
\left(1+ \left[\frac{4D-5}{12(2D-3)}-\frac{D^2+7D-6}{3D(2D-3)}\frac{\zeta(D-1)}{\zeta(D+1)}\right]\frac{d}{R}+\ldots \right).
\end{split}\end{equation}
It is easy to check that the respective leading terms coincide with the result of proximity force approximation (see Section \ref{sec5}). Hence, we can write
\begin{equation*}
\begin{split}
\frac{E_{\text{Cas}}^{\text{XY}}}{E_{\text{Cas}}^{\text{PFA},\text{XY}}}=\left\{1+\vartheta^{\text{XY}}\frac{d}{R}+o\left(\frac{d}{R}\right)\right\},
\end{split}
\end{equation*}
where
\begin{equation*}
\begin{split}
E_{\text{Cas}}^{\text{PFA},\text{DD}}=E_{\text{Cas}}^{\text{PFA},\text{NN}}=&-\frac{\hbar c \Gamma\left(D-\frac{1}{2}\right)\zeta(D+1)H\sqrt{R}}{2^{2D-\frac{1}{2}} \pi^{\frac{D-1}{2}}\Gamma\left(\frac{D}{2}\right)d^{D-\frac{1}{2}}},
\\
E_{\text{Cas}}^{\text{PFA},\text{DN}}=E_{\text{Cas}}^{\text{PFA},\text{ND}}=&(1-2^{-D})\frac{\hbar c \Gamma\left(D-\frac{1}{2}\right)\zeta(D+1)H\sqrt{R}}{2^{2D-\frac{1}{2}} \pi^{\frac{D-1}{2}}\Gamma\left(\frac{D}{2}\right)d^{D-\frac{1}{2}}},
\end{split}
\end{equation*} $$\vartheta^{\text{XY}}=\varkappa^{\text{XY}},$$
\begin{equation}\label{eq5_14_4}
\begin{split}
\varkappa^{\text{DD}}= &\frac{4D-5}{12(2D-3)}-\frac{(D-2)(D-3)}{3D(2D-3)} \frac{\zeta(D-1)}{\zeta(D+1)},\\
\varkappa^{\text{DN}}=&\frac{4D-5}{12(2D-3)}-\frac{(D-2)(D-3)}{3D(2D-3)}\frac{2^D-4}{2^D-1} \frac{\zeta(D-1)}{\zeta(D+1)},\\
\varkappa^{\text{ND}}= &\frac{4D-5}{12(2D-3)}-\frac{D^2+7D-6}{3D(2D-3)}\frac{2^D-4}{2^D-1}\frac{\zeta(D-1)}{\zeta(D+1)},\\
\varkappa^{\text{NN}}=&\frac{4D-5}{12(2D-3)}-\frac{D^2+7D-6}{3D(2D-3)}\frac{\zeta(D-1)}{\zeta(D+1)}.
\end{split}
\end{equation}The values of $\varkappa^{\text{XY}}$ are tabulated in Table \ref{t1} of Appendix \ref{a1} for $3\leq D\leq 6$.
$\vartheta$ measures the correction to the proximity force approximation. The dependence of $\vartheta$ on $D$ is plotted in Fig. \ref{f1}.

One observes some interesting phenomena. $\vartheta$ is positive when the cylinder is imposed with Dirichlet boundary conditions, which indicates a positive correction to the proximity force approximation; and $\vartheta$ is negative when the cylinder is imposed with Neumann boundary conditions, which indicates a negative correction. The correction is larger in the latter case. However, for all combinations of boundary conditions, we find that the magnitude of the correction decreases with dimension $D$.  In fact, from \eqref{eq5_14_4}, we find that when $D\gg 1$,
\begin{equation}\label{eq5_15_5}
\begin{split}
\vartheta^{\text{DY}}\sim \frac{5}{8}\frac{1}{D},\hspace{1cm} \vartheta^{\text{NY}}\sim -\frac{11}{8}\frac{1}{D},
\end{split}
\end{equation}which is inversely proportional to $D$. This is in big contrast to the sphere-plate interaction \cite{41}, where it is found that $\vartheta\sim -D/4$ regardless of the boundary conditions, and thus becomes negative when $D$ is large. In the sphere-plate case, the correction to the proximity force approximation becomes large when the dimension of spacetime is increased. However, for the cylinder-plate case, the correction to the proximity force approximation becomes smaller when the dimension of spacetime is increased.

\subsection{The case where one cylinder lies parallelly inside the other}\label{sec4_2}
In this case,
\begin{equation*}
\begin{split}
M_{n_i,n_{i+1}}^{\text{XY}}=T_1^{n_i,X}\sum_{n_i'=-\infty}^{\infty} I_{n_i'-n_i}(\gamma L)\widetilde{T}_2^{n_i',Y}I_{n_i'-n_{i+1}}(\gamma L).
\end{split}
\end{equation*}
Define
\begin{gather*}
a=\frac{R_1}{R_2-R_1},\quad b=\frac{R_2}{R_2-R_1},\quad \vep=\frac{d}{R_2-R_1},\\
n=n_0,\quad \omega=\gamma (R_2-R_1),
\end{gather*}
and make a change of variables
\begin{gather*}
  n_i=n+\tilde{n}_i,\quad 1\leq i\leq s,\\
n_i'=\frac{b}{a}n+\frac{b}{2a}\left(\tilde{n}_i+\tilde{n}_{i+1}\right)+q_i,\\
\omega=\frac{n\sqrt{1-\tau^2}}{a\tau}.
\end{gather*}Approximating  summations by integrations, we find that
\begin{equation}\begin{split}
E_{\text{Cas}}\sim &-\frac{\hbar c H}{2^{D-2}\pi^{\frac{D-1}{2}}\Gamma\left(\frac{D-1}{2}\right)R_1^{D-1}}\sum_{s=0}^{\infty}\frac{1}{s+1}  \int_{0}^{\infty}
dn \, n^{D-1}\int_0^1 d\tau\frac{(1-\tau^2)^{\frac{D-3}{2}}}{\tau^{D}}\int_{-\infty}^{\infty} d\tilde{n}_1\ldots\int_{-\infty}^{\infty} d\tilde{n}_s
M_{n_0,n_1}\ldots M_{n_{s},n_0},
\end{split}\end{equation}
where
\begin{equation*}
\begin{split}
M_{n_i,n_{i+1}}^{\text{XY}}=T_1^{n_i,X}\sum_{n_i'=-\infty}^{\infty} I_{n_i'-n_i}\left(\omega(1-\vep)\right)\widetilde{T}_2^{n_i',Y}I_{n_i'-n_{i+1}}\left(\omega(1-\vep)\right),
\end{split}
\end{equation*}
\begin{equation}
\begin{split}
T_1^{n_i, \text{D}} =&\frac{I_{n_i}(a\omega)}{K_{n_i}(a\omega)},\hspace{1cm}
T_1^{n_i, \text{N}}=\frac{I_{n_i}'(a\omega)}{K_{n_i}'(a\omega)},\\
\widetilde{T}_2^{n_i', \text{D}} =&\frac{K_{n_i'}(b\omega)}{I_{n_i'}(b\omega)},\hspace{1cm}
\widetilde{T}_2^{n_i', \text{N}} =\frac{K_{n_i'}'(b\omega)}{I_{n_i'}'(b\omega)}.\end{split}
\end{equation}
Using Debye asymptotic behavior of modified Bessel functions \eqref{eq3_26_6}, we find that
\begin{equation}
\begin{split}
M_{n_i,n_{i+1}}\sim & (-1)^{\alpha_{\text{X}}+\alpha_{\text{Y}}}\int_{-\infty}^{\infty} dq_i\frac{1}{2\pi\sqrt{\nu_2\nu_4}} \frac{e^{2\nu_1\eta(z_1)+\nu_2\eta(z_2)-2\nu_3\eta(z_3)+\nu_4\eta(z_4)}}{(1+z_2^2)^{\frac{1}{4}}(1+z_4^2)^{\frac{1}{4}}}\left(1+\mathcal{A}_2^{X}+\mathcal{B}_2^{Y}+\mathcal{C}_2 \right),
\end{split}
\end{equation}where
\begin{gather*}
\nu_1=n_i,\quad \nu_2=n_i'-n_i,\quad\nu_3=n_i',\quad \nu_4=n_i'-n_{i+1},\\
z_1=\frac{a\omega}{\nu_1},\quad z_2=\frac{\omega(1-\vep)}{\nu_2},\quad z_3=\frac{b\omega}{\nu_3},\quad z_4=\frac{\omega(1-\vep)}{\nu_4};
\end{gather*}
\begin{equation}
\begin{split}
\mathcal{A}_2^{\text{D}}=&\frac{2}{n}u_1(\tau),\hspace{1.3cm}
\mathcal{A}_2^{\text{N}}=\frac{2}{n}v_1(\tau),\\
\mathcal{B}_2^{\text{D}}=&-\frac{2a}{bn}u_1(\tau),\hspace{0.8cm}
\mathcal{B}_2^{\text{N}}=-\frac{2a}{bn}v_1(\tau),\\
\mathcal{C}_2=&\frac{2a}{n}u_1(\tau).
\end{split}
\end{equation}$\mathcal{A}_2$, $\mathcal{B}_2$ and $\mathcal{C}_2$ are terms of order $\vep$.
As in the cylinder-plate case, expanding each term keeping in mind that $n$ has order $\vep^{-1}$, and $\tilde{n}_i$ and $q_i$ has order $\vep^{-\frac{1}{2}}$, we obtain
\begin{equation}
\begin{split}
M_{n_i,n_{i+1}} \sim &C^{i_1-i_2}(-1)^{\alpha_{\text{X}}+\alpha_{\text{Y}}}\int_{-\infty}^{\infty} dq_i\frac{a\tau }{2\pi n }
\left(1+\mathcal{D}_{i,1}+\mathcal{D}_{i,2}\right)\exp\left(-\frac{2\vep n}{a\tau}-\frac{b\tau}{4n}(n_i-n_{i+1})^2-\frac{a^2\tau}{bn}q_i^2 \right)\\&\times\left(1+\mathcal{A}_2^{X}+\mathcal{B}_2^{Y}+\mathcal{C}_2 \right),\end{split}
\end{equation}where $\mathcal{D}_{i,1}$ and $\mathcal{D}_{i,2}$ are respectively terms of order $\sqrt{\vep}$ and $\vep$. The integration over $q_i$ is straightforward and gives an expansion of the form
\begin{equation}
\begin{split}
M_{n_i,n_{i+1}} \sim &C^{i_1-i_2}(-1)^{\alpha_{\text{X}}+\alpha_{\text{Y}}} \frac{\sqrt{b\tau} }{2\sqrt{\pi  n} }
\left(1+\mathcal{G}_{i,1}+\mathcal{G}_{i,2}\right)\exp\left(-\frac{2\vep n}{a\tau}-\frac{b\tau}{4n}(n_i-n_{i+1})^2 \right) \left(1+\mathcal{A}_2^{X}+\mathcal{B}_2^{Y}+\mathcal{C}_2 \right).
\end{split}
\end{equation}
The rest is similar to the cylinder-plate case. We find that the up to the next-to-leading order term, the Casimir interaction energy can be written as
\begin{equation*}
\begin{split}
E_{\text{Cas}}^{\text{DD}}=&E_{\text{Cas}}^{\text{DD}, \text{PFA}}\left(1+\frac{4D-5}{4(2D-3)}\frac{d}{R_2-R_1}+\varkappa^{\text{DD}} \frac{d}{R_1}-\varkappa^{\text{DD}}\frac{d}{R_2}\right),\\
E_{\text{Cas}}^{\text{DN}}=&E_{\text{Cas}}^{\text{DN}, \text{PFA}}\left(1+\frac{4D-5}{4(2D-3)}\frac{d}{R_2-R_1}+\varkappa^{\text{DN}} \frac{d}{R_1}-\varkappa^{\text{ND}}\frac{d}{R_2}\right),\\
E_{\text{Cas}}^{\text{ND}}=&E_{\text{Cas}}^{\text{ND}, \text{PFA}}\left(1+\frac{4D-5}{4(2D-3)}\frac{d}{R_2-R_1}+\varkappa^{\text{ND}} \frac{d}{R_1}-\varkappa^{\text{DN}}\frac{d}{R_2}\right),\\
E_{\text{Cas}}^{\text{NN}}=&E_{\text{Cas}}^{\text{NN}, \text{PFA}}\left(1+\frac{4D-5}{4(2D-3)}\frac{d}{R_2-R_1}+\varkappa^{\text{NN}} \frac{d}{R_1}-\varkappa^{\text{NN}}\frac{d}{R_2}\right).
\end{split}
\end{equation*}Here, $\varkappa^{\text{XY}}$ is defined in \eqref{eq5_14_4}, and are equal to the $\vartheta^{\text{XY}}$ for the cylinder-plate case, and $E_{\text{Cas}}^{\text{XY}, \text{PFA}}$ is the leading term that coincides with the proximity force approximation. They are given explicitly by
\begin{equation*}
\begin{split}
E_{\text{Cas}, \text{PFA}}^{\text{DD}}=E_{\text{Cas}, \text{PFA}}^{\text{NN}}=&-\frac{\hbar c H\Gamma\left(D-\frac{1}{2}\right)\zeta(D+1)}{2^{2D-\frac{1}{2}} \pi^{\frac{D-1}{2}}\Gamma\left(\frac{D}{2}\right)d^{D-\frac{1}{2}}}\sqrt{\frac{R_1R_2}{R_2-R_1}},\\
E_{\text{Cas}, \text{PFA}}^{\text{DN}}=E_{\text{Cas}, \text{PFA}}^{\text{ND}}=&(1-2^{-D})\frac{\hbar c H\Gamma\left(D-\frac{1}{2}\right)\zeta(D+1)}{2^{2D-\frac{1}{2}} \pi^{\frac{D-1}{2}}\Gamma\left(\frac{D}{2}\right)d^{D-\frac{1}{2}}}\sqrt{\frac{R_1R_2}{R_2-R_1}}.
\end{split}
\end{equation*}
Hence, we have
\begin{equation*}
\begin{split}
\frac{E_{\text{Cas}}^{\text{XY}}}{E_{\text{Cas}}^{\text{PFA},\text{XY}}}=\left\{1+\vartheta^{\text{XY}}\frac{d}{R_2-R_1}+o\left(\frac{d}{R_2-R_1}\right)\right\},
\end{split}
\end{equation*}
where
\begin{equation}\label{eq5_15_6}
\begin{split}
\vartheta^{\text{DD}}=&\frac{4D-5}{4(2D-3)}+\varkappa^{\text{DD}}\frac{1}{a}- \varkappa^{\text{DD}}\frac{1}{b},\\
\vartheta^{\text{DN}}=&\frac{4D-5}{4(2D-3)}+\varkappa^{\text{DN}}\frac{1}{a}- \varkappa^{\text{ND}}\frac{1}{b},\\
\vartheta^{\text{ND}}=&\frac{4D-5}{4(2D-3)}+\varkappa^{\text{ND}}\frac{1}{a}- \varkappa^{\text{DN}}\frac{1}{b},\\
\vartheta^{\text{NN}}=&\frac{4D-5}{4(2D-3)}+\varkappa^{\text{NN}}\frac{1}{a}- \varkappa^{\text{NN}}\frac{1}{b}.
\end{split}
\end{equation}

\begin{figure}[h]
\epsfxsize=0.4\linewidth \epsffile{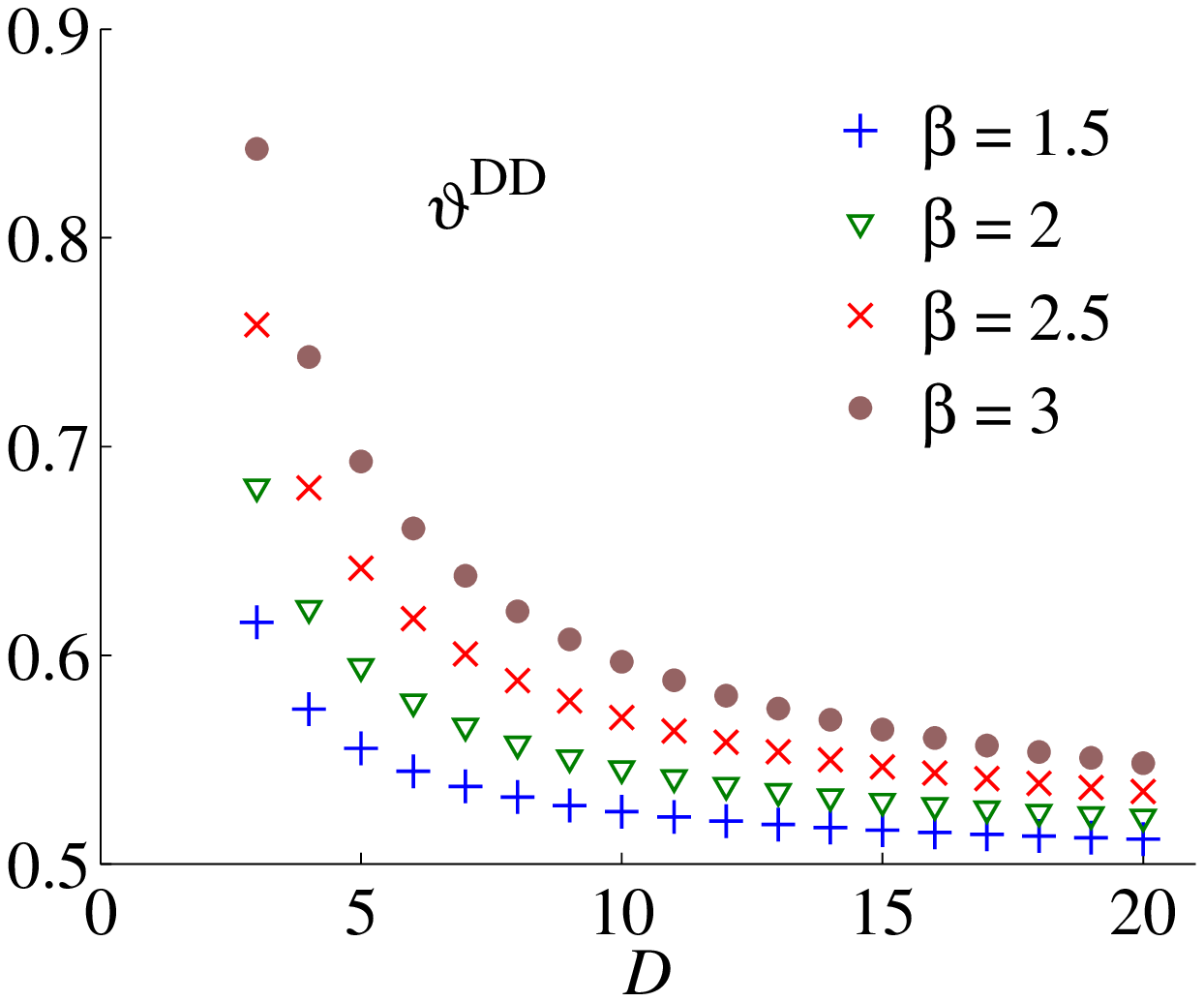} \epsfxsize=0.4\linewidth \epsffile{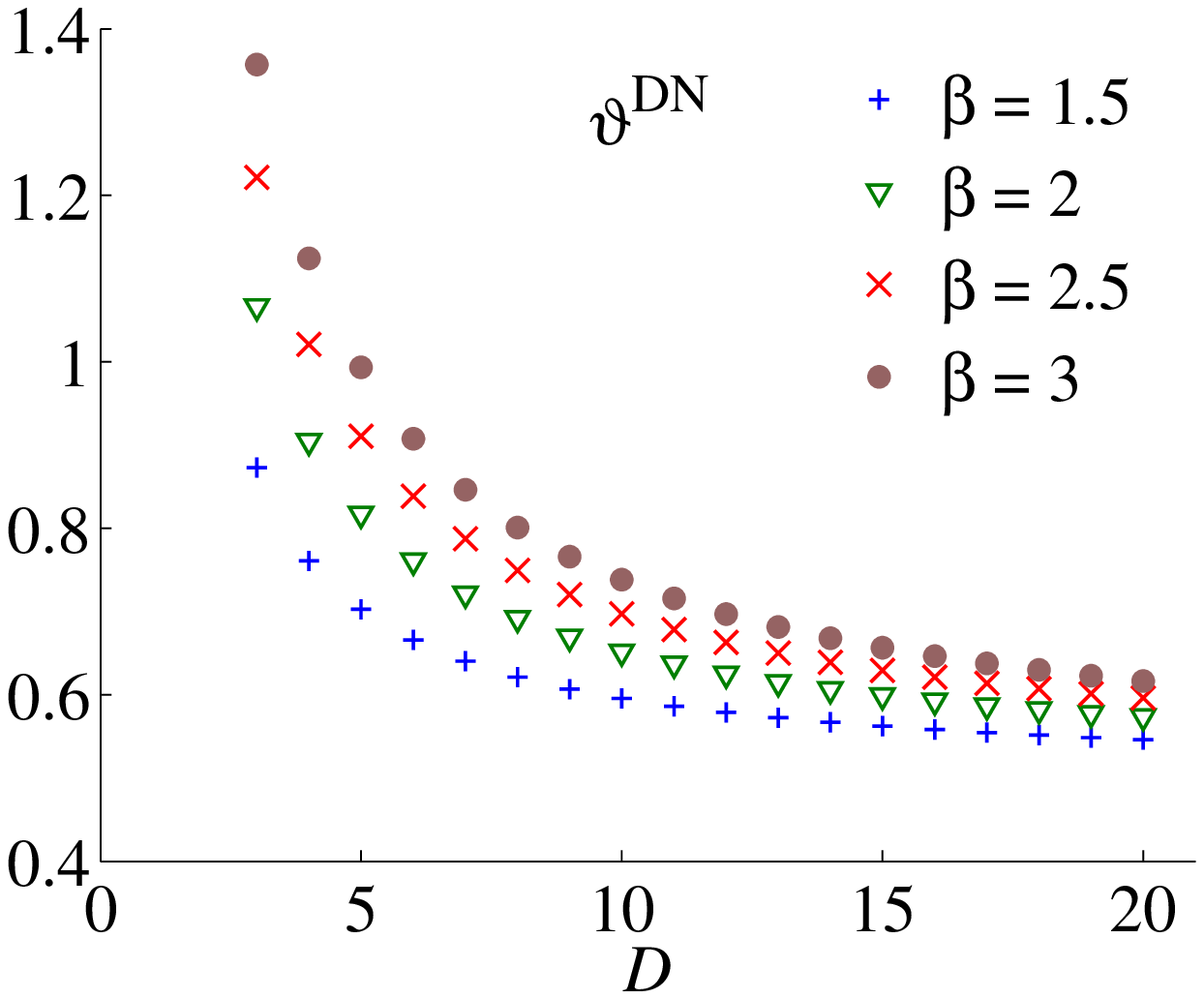}\\\epsfxsize=0.4\linewidth \epsffile{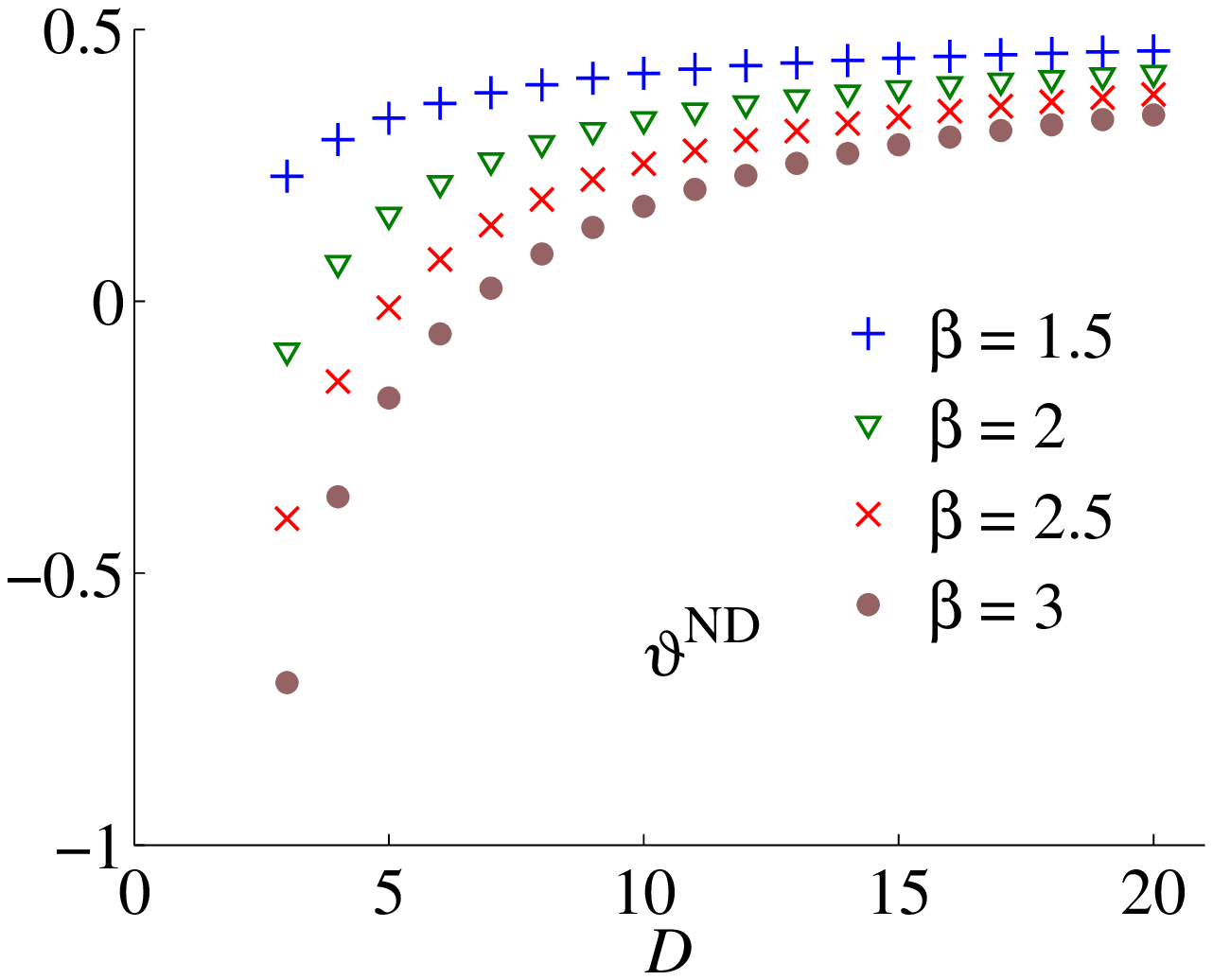} \epsfxsize=0.4\linewidth \epsffile{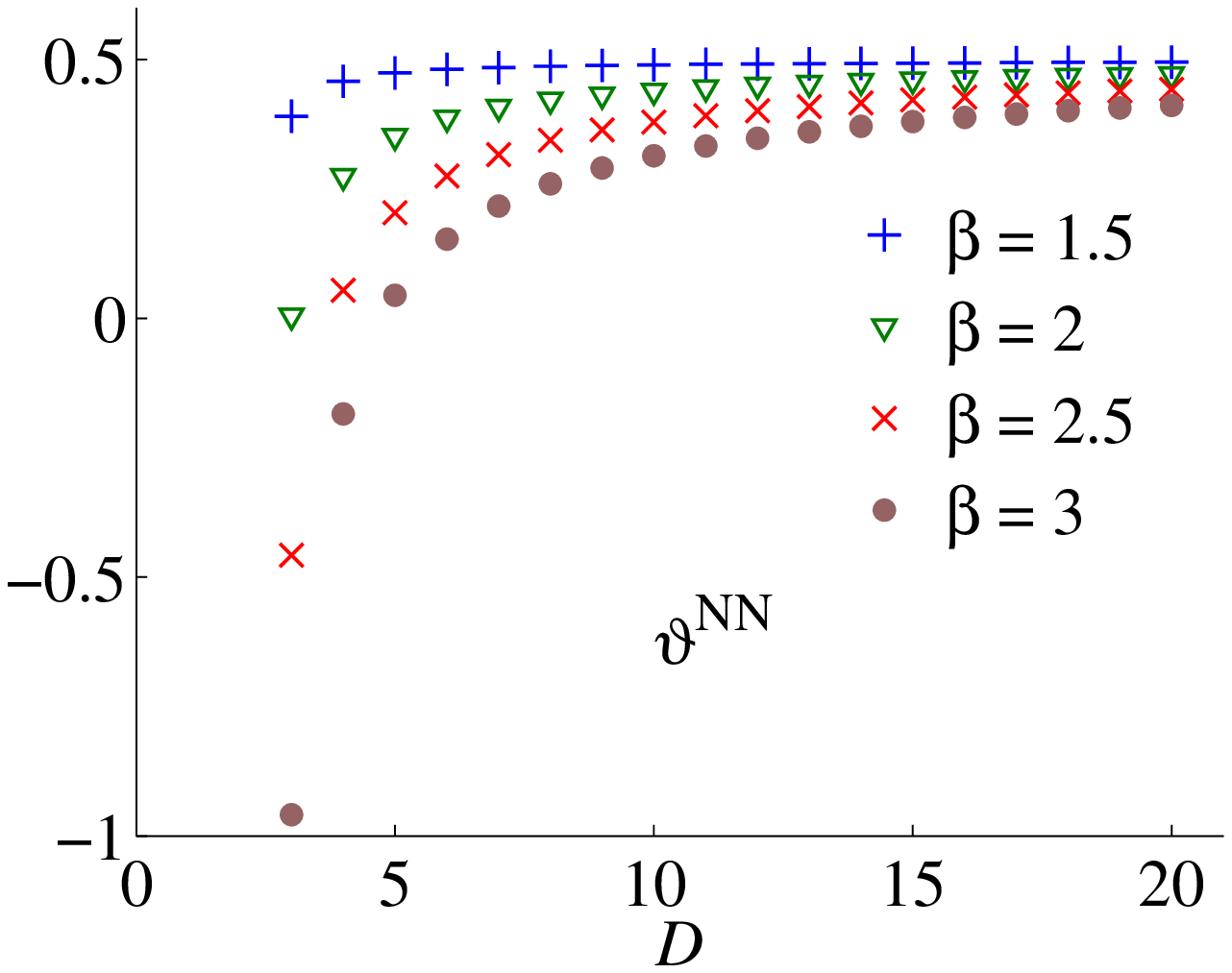}
 \caption{\label{f2} The dependence of $\vartheta$  on dimension $D$ and radii ratio $\beta$ for different combinations of boundary conditions.}\end{figure}

Recall that $b=a+1$. Hence, we can regard $\vartheta$ as depending on dimension $D$ and $\beta=b/a=R_2/R_1$--the ratio of the radius of the larger cylinder to the radius of the smaller cylinder. Then
\begin{equation*}
a=\frac{1}{\beta-1},\hspace{1cm}b=\frac{\beta}{\beta-1}.
\end{equation*}
In Fig. \ref{f2}, we plot the dependence of $\vartheta$ on the dimension $D$ and radii ratio $\beta$ for different boundary conditions.

We observe that $\vartheta$ is always positive when the inner cylinder is imposed with Dirichlet boundary conditions. When the inner cylinder is imposed with Neumann boundary conditions, $\vartheta$ can be positive or negative depending on the dimension $D$ and the ratio of the radii of the cylinders.
When $D$ is large, we observe some universal behavior. In fact, from \eqref{eq5_15_6} and the asymptotic behavior of $\varkappa^{\text{XY}}$ obtained in \eqref{eq5_15_5}, we find that when $D\gg 1$,
\begin{equation*}
\begin{split}
\vartheta^{\text{XY}}\sim & \frac{1}{2},
\end{split}
\end{equation*}regardless of the boundary conditions. This agrees with the graphs we obtained in Figure \ref{f2}. The dominating term actually comes from
$$\frac{4D-5}{4(2D-3)},$$ which is universal for all boundary conditions.

\subsection{The case where two parallel cylinders are exterior to each other}\label{sec4_3}
In this case,
\begin{equation*}
\begin{split}
M_{n_i,n_{i+1}}^{\text{XY}}=T_1^{n_i,X}\sum_{n_i'=-\infty}^{\infty} K_{n_i'+n_i}(\gamma L)T_2^{n_i',Y}K_{n_i'+n_{i+1}}(\gamma L).
\end{split}
\end{equation*}
Define
\begin{gather*}
a=\frac{R_1}{R_1+R_2},\quad b=\frac{R_2}{R_1+R_2},\quad \vep=\frac{d}{R_1+R_2},\\
n=n_0,\quad \omega=\gamma (R_1+R_2),
\end{gather*}
and make a change of variables
\begin{gather*}
  n_i=n+\tilde{n}_i,\quad 1\leq i\leq s,\\
n_i'=\frac{b}{a}n+\frac{b}{2a}\left(\tilde{n}_i+\tilde{n}_{i+1}\right)+q_i,\\
\omega=\frac{n\sqrt{1-\tau^2}}{a\tau}.
\end{gather*}As in the previous case, we find that
\begin{equation}\begin{split}
E_{\text{Cas}}\sim &-\frac{\hbar c H}{2^{D-2}\pi^{\frac{D-1}{2}}\Gamma\left(\frac{D-1}{2}\right)R_1^{D-1}}\sum_{s=0}^{\infty}\frac{1}{s+1}  \int_{0}^{\infty}
dn \, n^{D-1}\int_0^1 d\tau\frac{(1-\tau^2)^{\frac{D-3}{2}}}{\tau^{D}}\int_{-\infty}^{\infty} d\tilde{n}_1\ldots\int_{-\infty}^{\infty} d\tilde{n}_s
M_{n_0,n_1}\ldots M_{n_{s},n_0},
\end{split}\end{equation}
where
\begin{equation*}
\begin{split}
M_{n_i,n_{i+1}}^{\text{XY}}=T_1^{n_i,X}\sum_{n_i'=-\infty}^{\infty} K_{n_i'+n_i}\left(\omega(1+\vep)\right)T_2^{n_i',Y}K_{n_i'+n_{i+1}}\left(\omega(1+\vep)\right),
\end{split}
\end{equation*}
\begin{equation}
\begin{split}
T_1^{n_i, \text{D}} =&\frac{I_{n_i}(a\omega)}{K_{n_i}(a\omega)},\hspace{1cm}
T_1^{n_i, \text{N}}=\frac{I_{n_i}'(a\omega)}{K_{n_i}'(a\omega)},\\
\widetilde{T}_2^{n_i', \text{D}} =&\frac{I_{n_i'}(b\omega)}{K_{n_i'}(b\omega)},\hspace{1cm}
\widetilde{T}_2^{n_i', \text{N}} =\frac{I_{n_i'}'(b\omega)}{K_{n_i'}'(b\omega)}.\end{split}
\end{equation}
Using Debye asymptotic behavior of modified Bessel functions \eqref{eq3_26_6}, we find that
\begin{equation}
\begin{split}
M_{n_i,n_{i+1}}\sim & (-1)^{\alpha_{\text{X}}+\alpha_{\text{Y}}}\int_{-\infty}^{\infty} dq_i\frac{1}{2\pi\sqrt{\nu_2\nu_4}} \frac{e^{2\nu_1\eta(z_1)-\nu_2\eta(z_2)+2\nu_3\eta(z_3)-\nu_4\eta(z_4)}}{(1+z_2^2)^{\frac{1}{4}}(1+z_4^2)^{\frac{1}{4}}}\left(1+\mathcal{A}_2^{X}+\mathcal{B}_2^{Y}+\mathcal{C}_2 \right),
\end{split}
\end{equation}where
\begin{gather*}
\nu_1=n_i,\quad \nu_2=n_i'+n_i,\quad\nu_3=n_i',\quad \nu_4=n_i'+n_{i+1},\\
z_1=\frac{a\omega}{\nu_1},\quad z_2=\frac{\omega(1+\vep)}{\nu_2},\quad z_3=\frac{b\omega}{\nu_3},\quad z_4=\frac{\omega(1+\vep)}{\nu_4};
\end{gather*}
\begin{equation}
\begin{split}
\mathcal{A}_2^{\text{D}}=&\frac{2}{n}u_1(\tau),\hspace{1cm}
\mathcal{A}_2^{\text{N}}=\frac{2}{n}v_1(\tau),\\
\mathcal{B}_2^{\text{D}}=&\frac{2a}{bn}u_1(\tau),\hspace{1cm}
\mathcal{B}_2^{\text{N}}=\frac{2a}{bn}v_1(\tau),\\
\mathcal{C}_2=&-\frac{2a}{n}u_1(\tau).
\end{split}
\end{equation}
As  before, expanding each term according to orders of $\vep$ gives
\begin{equation}
\begin{split}
M_{n_i,n_{i+1}} \sim &C^{i_1-i_2}(-1)^{\alpha_{\text{X}}+\alpha_{\text{Y}}}\int_{-\infty}^{\infty} dq_i\frac{a\tau }{2\pi n }
\left(1+\mathcal{D}_{i,1}+\mathcal{D}_{i,2}\right)\exp\left(-\frac{2\vep n}{a\tau}-\frac{b\tau}{4n}(n_i-n_{i+1})^2-\frac{a^2\tau}{bn}q_i^2 \right)\\&\times\left(1+\mathcal{A}_2^{X}+\mathcal{B}_2^{Y}+\mathcal{C}_2 \right),\end{split}
\end{equation}where $\mathcal{D}_{i,1}$ and $\mathcal{D}_{i,2}$ are respectively terms of order $\sqrt{\vep}$ and $\vep$.
The rest is similar to the case where one cylinder is inside the other. We find that the up to next-to-leading order term, the Casimir interaction energy can be written as
\begin{equation}\label{eq5_18_2}
\begin{split}
E_{\text{Cas}}^{\text{DD}}=&E_{\text{Cas}}^{\text{DD}, \text{PFA}}\left(1-\frac{4D-5}{4(2D-3)}\frac{d}{R_1+R_2}+\varkappa^{\text{DD}} \frac{d}{R_1}+\varkappa^{\text{DD}}\frac{d}{R_2}\right),\\
E_{\text{Cas}}^{\text{DN}}=&E_{\text{Cas}}^{\text{DN}, \text{PFA}}\left(1-\frac{4D-5}{4(2D-3)}\frac{d}{R_1+R_2}+\varkappa^{\text{DN}} \frac{d}{R_1}+\varkappa^{\text{ND}}\frac{d}{R_2}\right),\\
E_{\text{Cas}}^{\text{ND}}=&E_{\text{Cas}}^{\text{ND}, \text{PFA}}\left(1-\frac{4D-5}{4(2D-3)}\frac{d}{R_1+R_2}+\varkappa^{\text{ND}} \frac{d}{R_1}+\varkappa^{\text{DN}}\frac{d}{R_2}\right),\\
E_{\text{Cas}}^{\text{NN}}=&E_{\text{Cas}}^{\text{NN}, \text{PFA}}\left(1-\frac{4D-5}{4(2D-3)}\frac{d}{R_1+R_2}+\varkappa^{\text{NN}} \frac{d}{R_1}+\varkappa^{\text{NN}}\frac{d}{R_2}\right).
\end{split}
\end{equation}Here,  $E_{\text{Cas}}^{\text{XY}, \text{PFA}}$ is the leading term that coincides with the proximity force approximation. They are given explicitly by
\begin{equation*}
\begin{split}
E_{\text{Cas}, \text{PFA}}^{\text{DD}}=E_{\text{Cas}, \text{PFA}}^{\text{NN}}=&-\frac{\hbar c H\Gamma\left(D-\frac{1}{2}\right)\zeta(D+1)}{2^{2D-\frac{1}{2}} \pi^{\frac{D-1}{2}}\Gamma\left(\frac{D}{2}\right)d^{D-\frac{1}{2}}}\sqrt{\frac{R_1R_2}{R_1+R_2}},\\
E_{\text{Cas}, \text{PFA}}^{\text{DN}}=E_{\text{Cas}, \text{PFA}}^{\text{ND}}=&(1-2^{-D})\frac{\hbar c H\Gamma\left(D-\frac{1}{2}\right)\zeta(D+1)}{2^{2D-\frac{1}{2}} \pi^{\frac{D-1}{2}}\Gamma\left(\frac{D}{2}\right)d^{D-\frac{1}{2}}}\sqrt{\frac{R_1R_2}{R_1+R_2}}.
\end{split}
\end{equation*}
Hence, we have
\begin{equation*}
\begin{split}
\frac{E_{\text{Cas}}^{\text{XY}}}{E_{\text{Cas}}^{\text{PFA},\text{XY}}}=\left\{1+\vartheta^{\text{XY}}\frac{d}{R_1+R_2}+o\left(\frac{d}{R_1+R_2}\right)\right\},
\end{split}
\end{equation*}
where
\begin{equation}\label{eq5_15_7}
\begin{split}
\vartheta^{\text{DD}}=&-\frac{4D-5}{4(2D-3)}+\varkappa^{\text{DD}}\frac{1}{a}+ \varkappa^{\text{DD}}\frac{1}{b},\\
\vartheta^{\text{DN}}=&-\frac{4D-5}{4(2D-3)}+\varkappa^{\text{DN}}\frac{1}{a}+ \varkappa^{\text{ND}}\frac{1}{b},\\
\vartheta^{\text{ND}}=&-\frac{4D-5}{4(2D-3)}+\varkappa^{\text{ND}}\frac{1}{a}+ \varkappa^{\text{DN}}\frac{1}{b},\\
\vartheta^{\text{NN}}=&-\frac{4D-5}{4(2D-3)}+\varkappa^{\text{NN}}\frac{1}{a}+ \varkappa^{\text{NN}}\frac{1}{b}.
\end{split}
\end{equation}

\begin{figure}[h]
\epsfxsize=0.4\linewidth \epsffile{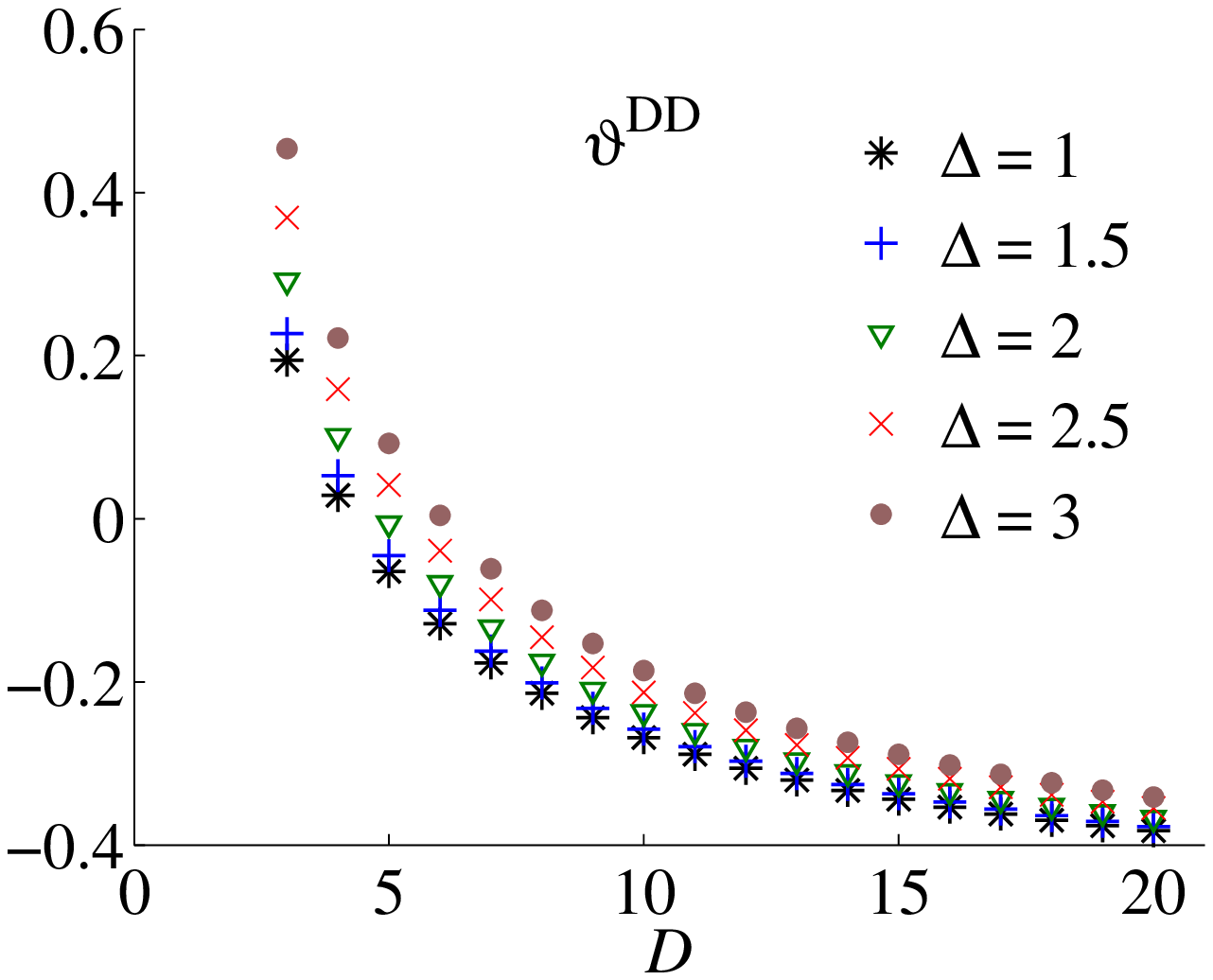} \epsfxsize=0.4\linewidth \epsffile{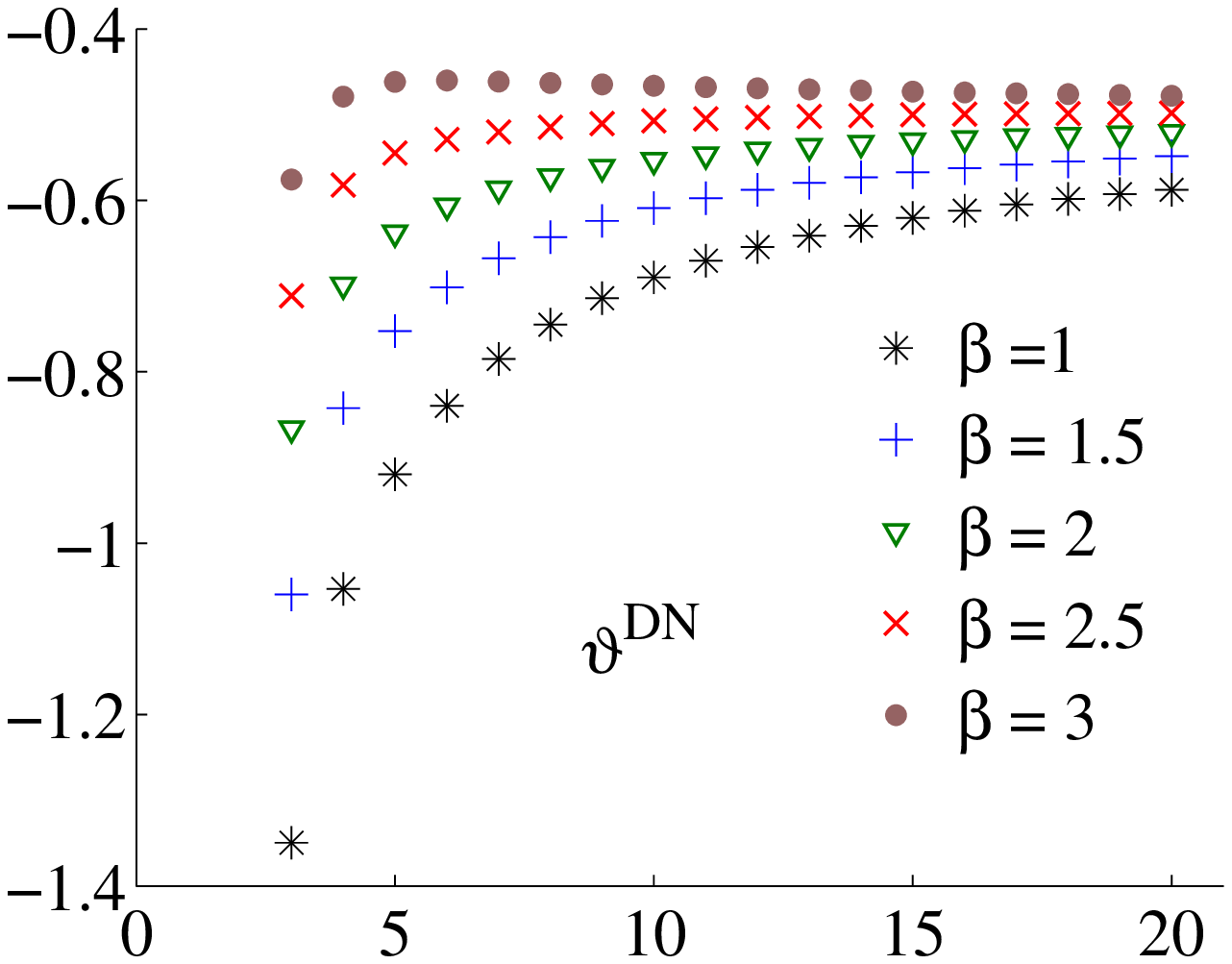}\\\epsfxsize=0.4\linewidth \epsffile{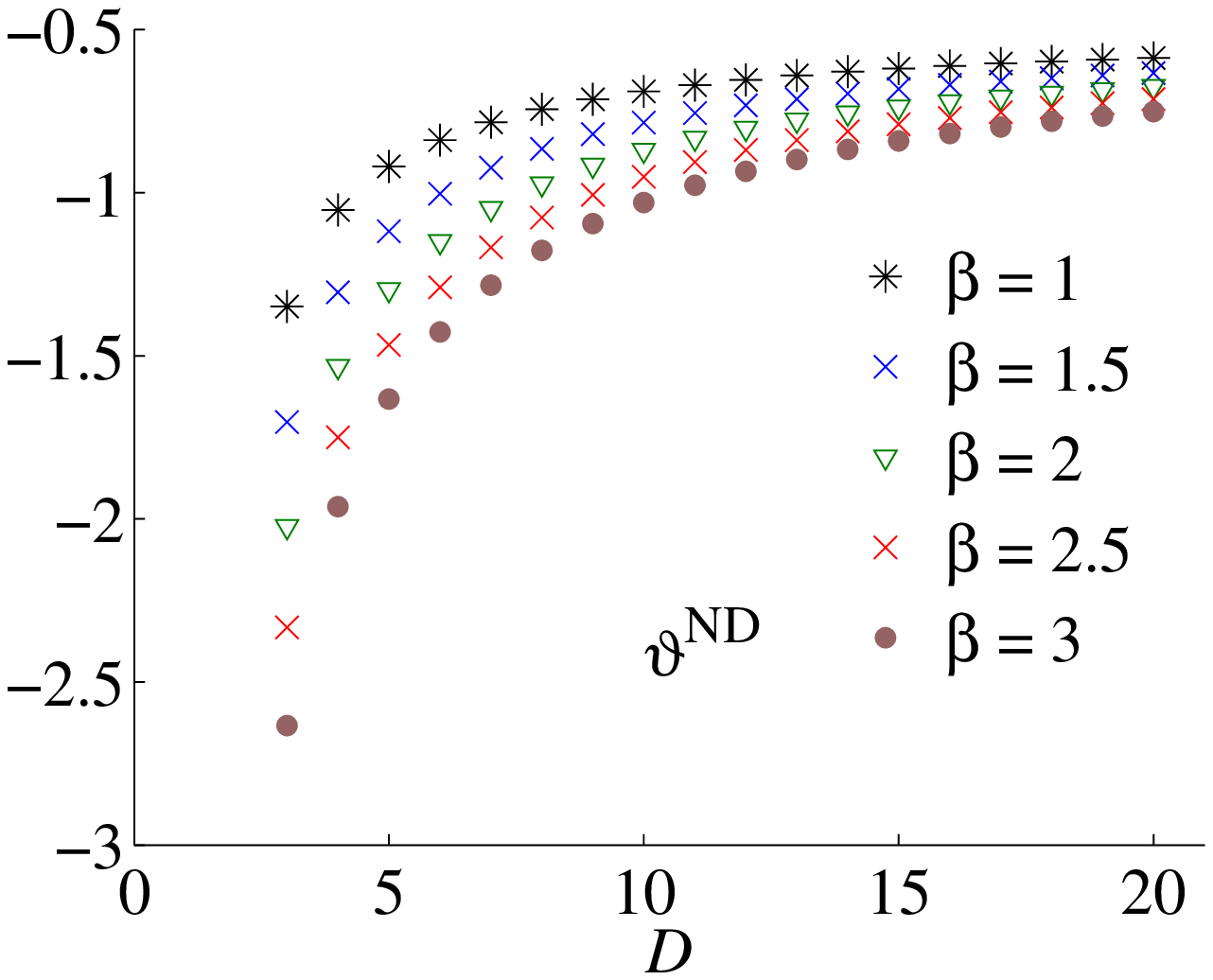} \epsfxsize=0.4\linewidth \epsffile{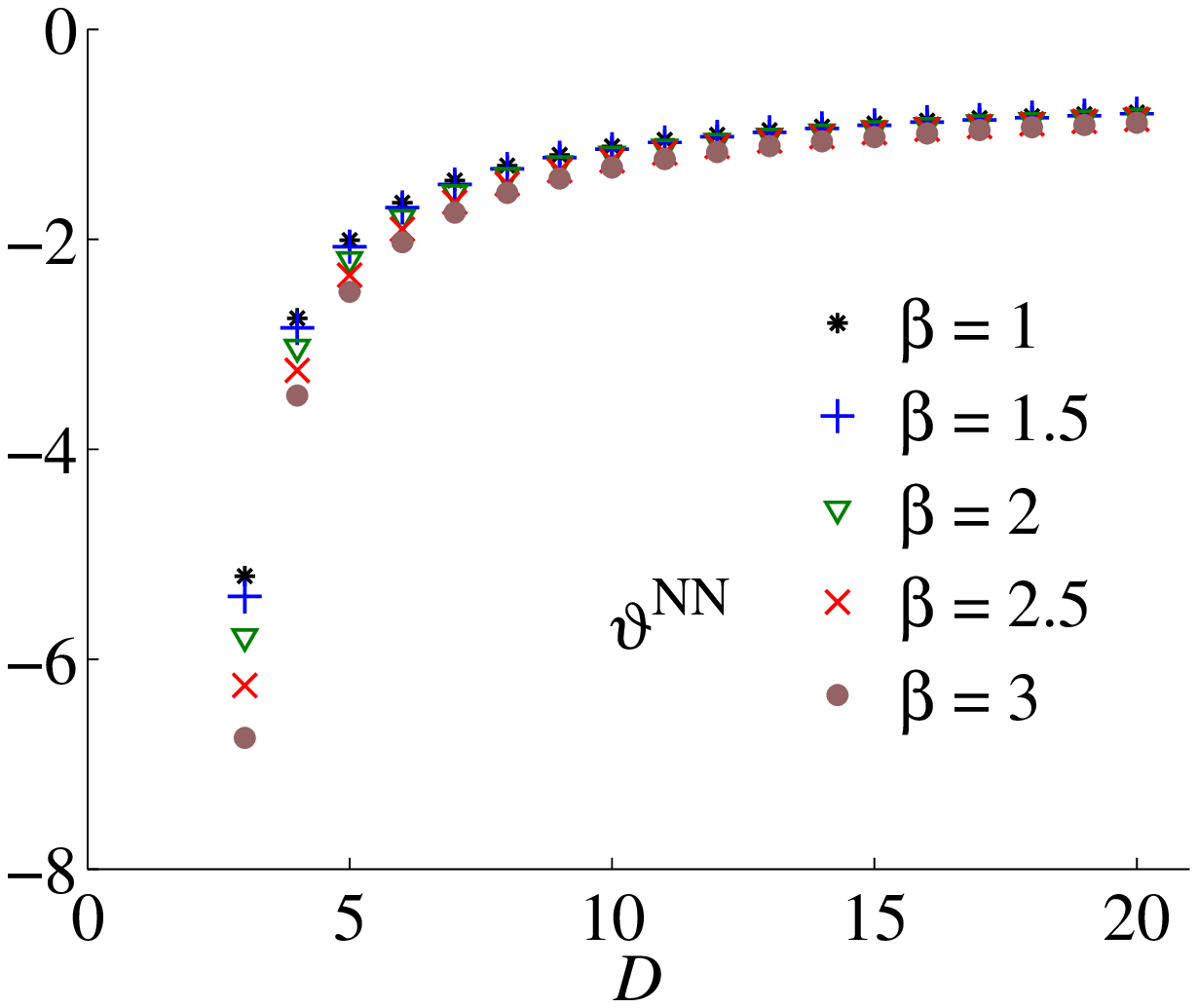}
 \caption{\label{f3} The dependence of $\vartheta$  on dimension $D$ and radii ratio $\Delta$ for different combinations of boundary conditions.}\end{figure}

Recall that $b=1-a$.  Hence, we can regard $\vartheta$ as depending on dimension $D$ and $\Delta=b/a=R_2/R_1$--the ratio of the radii of the cylinders. Without loss of generality, we can assume that $R_2\geq R_1$. Then $\Delta\geq 1$,
\begin{equation*}
a=\frac{1}{\Delta+1},\hspace{1cm}b=\frac{\Delta}{\Delta+1}.
\end{equation*}

In Fig. \ref{f3}, we plot the dependence of $\vartheta$ on the dimension $D$ and radii ratio $\Delta$ for different boundary conditions.

We observe that $\vartheta$ is always negative for DN, ND and NN boundary conditions. When both cylinders are imposed with Dirichlet boundary conditions, $\vartheta$ can be positive or negative depending on the dimension $D$ and the ratio of the radii of the cylinders.
When $D$ is large, we observe some universal behavior. In fact, from \eqref{eq5_15_7} and the asymptotic behavior of $\varkappa^{\text{XY}}$ obtained in \eqref{eq5_15_5}, we find that when $D\gg 1$,
\begin{equation*}
\begin{split}
\vartheta^{\text{XY}}\sim & -\frac{1}{2},
\end{split}
\end{equation*}regardless of the boundary conditions. This agrees with the graphs we obtained in Figure \ref{f3}. The dominating term actually comes from
$$-\frac{4D-5}{4(2D-3)},$$ which is universal for all boundary conditions.

\section{Postulate for derivative expansion formula}\label{sec5}
In a series of papers \cite{52,54,55}, Fosco,   Lombardo and  Mazzitelli used derivative expansion to compute the  Casimir interaction energy between a curved surface and a plane with Dirichlet boundary conditions up to the next-to-leading order term. In \cite{52}, they showed that  the derivative expansion of the   Casimir interaction energy  is given by
\begin{equation}\begin{split}
E_{\text{Cas}}^{\text{DE} }=\hbar c\int_{S} d^{D-1}\mathbf{x}_{\perp} \left(b_0(D)\frac{1}{\left|\psi(\mathbf{x}_{\perp})\right|^D}+b_2(D)\frac{\Vert\nabla\psi\Vert^2}{\left|\psi(\mathbf{x}_{\perp})\right|^D}+\ldots\right).
\end{split}\end{equation}Here $\mathbf{x}_{\perp}=(x_2,\ldots,x_D)$, $x_1=\psi(\mathbf{x}_{\perp}), \mathbf{x}_{\perp}\in S$ defines the position of the curved surface with respect to the plane at $x_1=0$, and
\begin{equation}\begin{split}
b_0(D)=&-\frac{\Gamma\left(\frac{D+1}{2}\right)\zeta(D+1)}{2^{D+1}\pi^{\frac{D+1}{2}}},\\
b_2(D)=&-\frac{1}{3\times 2^{D+2}\pi^{\frac{D+1}{2}}}\left\{-\frac{(D-3)(D-1)(D-2)}{2D} \Gamma\left(\frac{D-1}{2}\right)\zeta(D-1)+ (D+1)\Gamma\left(\frac{D+1}{2}\right)\zeta(D+1)\right\}\\
=&b_0(D)\left(\frac{D+1}{6}-\frac{(D-2)(D-3)}{6D}\frac{\zeta(D-1)}{\zeta(D+1)}\right).
\end{split}\end{equation}

Consider the cylinder-plate interaction. We can take the cylinder to be $(x_1-L)^2+x_2^2=R^2$. Then
\begin{equation*}\begin{split}
\psi(\mathbf{x}_{\perp})=&\psi(x_2)=L-\sqrt{R^2-x_2^2},\\
\nabla \psi =&\frac{x_2}{\sqrt{R^2-x_2^2}}\mathbf{e}_2.
\end{split}\end{equation*}Some computations give
\begin{equation*}\begin{split}
\int_{S} d^{D-1}\mathbf{x}_{\perp} \frac{1}{\left|\psi(\mathbf{x}_{\perp})\right|^D}=& 2H\int_{0}^R dx_2 \frac{1}{\left(L-\sqrt{R^2-x_2^2}\right)^D}\\
= &\frac{2 H \sqrt{R}}{\sqrt{2} d^{D-\frac{1}{2}} }\frac{\pi\Gamma\left(D-\frac{1}{2}\right)}{2^{D-1}\Gamma\left(\frac{D}{2}\right)\Gamma\left(\frac{D+1}{2}\right)}\left(1-\frac{3d}{4R}\frac{1}{2D-3}+\ldots\right),
\end{split}\end{equation*}
\begin{equation*}\begin{split}
\int_{S} d^{D-1}\mathbf{x}_{\perp} \frac{1}{\left|\psi(\mathbf{x}_{\perp})\right|^D}=& 2H\int_{0}^R dx_2 \frac{1}{\left(L-\sqrt{R^2-x_2^2}\right)^D}\frac{x_2^2}{R_2^2-x_2^2}\\
=&\frac{2 H \sqrt{R}}{\sqrt{2} d^{D-\frac{1}{2}} }\frac{\pi\Gamma\left(D-\frac{1}{2}\right)}{2^{D-1}\Gamma\left(\frac{D}{2}\right)\Gamma\left(\frac{D+1}{2}\right)} \frac{2}{2D-3}\frac{d}{R}+\ldots.
\end{split}\end{equation*}
Hence, derivative expansion gives
\begin{equation}\begin{split}
E_{\text{Cas}}^{\text{DE},\text{DD}}\sim & - \frac{\hbar c \Gamma\left(D-\frac{1}{2}\right)\zeta(D+1)H\sqrt{R}}{2^{2D-\frac{1}{2}}\pi^{\frac{D-1}{2}}\Gamma\left(\frac{D}{2}\right)d^{D-\frac{1}{2}}}
\left(1+ \left[\frac{4D-5}{12(2D-3)}-\frac{(D-2)(D-3)}{3D(2D-3)} \frac{\zeta(D-1)}{\zeta(D+1)}\right]\frac{d}{R}+\ldots \right),
\end{split}\end{equation}agreeing with the result we obtain in the first formula of \eqref{eq5_18_1}.

Encouraged by this, we would like to give a postulate for the result of derivative expansion for DN, ND and NN boundary conditions, in the case where the two interacting objects are both curved. Inspired by \cite{53}, let us formulate the following ansatz for the small separation asymptotic behavior of the Casimir interaction energy between two curved objects in $(D+1)$-dimensional Minkowski spacetime:
 \begin{equation}\label{eq3_27_5}\begin{split}
 E_{\text{Cas}}^{\text{DE}}=&\int_{\Sigma}d^{D-1}\mathbf{x}_{\perp}\, \mathcal{E}_{\text{Cas}}^{\parallel}(H)\Bigl(1+\beta_1(H)\nabla H_1\cdot\nabla H_1+\beta_2(H)\nabla H_2\cdot\nabla H_2+\beta_{\times}(H)\nabla H_1\cdot \nabla H_2+\ldots\Bigr),
 \end{split}\end{equation}where $\mathcal{E}_{\text{Cas}}^{\parallel}$ is the Casimir energy density between two parallel plates, $\Sigma$ can be taken to be the $x_1=0$ plane parametrized by $\mathbf{x}_{\perp}=(x_2,\ldots,x_D)$,   $x_1=H_1(\mathbf{x}_{\perp})$ and $x_2=H_2(\mathbf{x}_{\perp})$ are the height profiles of the two objects with respect to $\Sigma$, and $H=H_1-H_2$ is the height difference. When the second object is a plane we can take it as the plane $x_1=0$ and then $H_2=0$.

Notice that
\begin{equation}\begin{split}
\mathcal{E}_{\text{Cas}}^{\parallel, \text{XY}}(H)=\frac{b_0^{\text{XY}}(D)}{H^{D}},
\end{split}\end{equation}
where
\begin{equation}\begin{split}
b_0^{\text{DD}}(D)=b_0^{\text{NN}}(D)=&-\frac{\hbar c\Gamma\left(\frac{D+1}{2}\right)\zeta(D+1)}{2^{D+1}\pi^{\frac{D+1}{2}}},\\
b_0^{\text{DN}}(D)=b_0^{\text{ND}}(D)=&\left(1-2^{-D}\right)\frac{\hbar c\Gamma\left(\frac{D+1}{2}\right)\zeta(D+1)}{2^{D+1}\pi^{\frac{D+1}{2}}}.
\end{split}\end{equation}
According to the result of \cite{52} we mentioned above,
\begin{equation}\begin{split}
\beta_1^{\text{DD}}=\beta_2^{\text{DD}}=\frac{D+1}{6}-\frac{(D-2)(D-3)}{6D}\frac{\zeta(D-1)}{\zeta(D+1)}.
\end{split}\end{equation}We are going to determine the values of $\beta_1$, $\beta_2$ and $\beta_{\times}$ for DD, DN, ND and NN boundary conditions based on our results on the small separation asymptotic expansions of the Casimir interaction energies between two cylinders exterior to each other.

We can take the two cylinders to have  height profiles
$$H_1=L_1-\sqrt{R_1^2-x_2^2}, \hspace{1cm}H_2=-L_2+\sqrt{R_2^2-x_2^2}.$$ Then $L=L_1+L_2$ is the distance between the centers of the cylinders.
Assuming that $R_1<R_2$,  some tedious computations give
\begin{equation}\begin{split}
\int_{\Sigma} d^{D-1}\mathbf{x}_{\perp} \frac{1}{H^D}=&2H\int_{0}^{R_1}dx\frac{1}{\left(L-\sqrt{R_1^2-x^2}-\sqrt{R_2^2-x^2}\right)^D}\\
=&\frac{H}{d^{D-\frac{1}{2}}}\sqrt{ \frac{2\pi R_1R_2}{R_1+R_2} }\frac{\Gamma\left(D-\frac{1}{2}\right)}{\Gamma(D)}\left\{1+\frac{9}{4(2D-3)}\frac{d}{R_1+R_2}-\frac{3}{4(2D-3)}\left(\frac{d}{R_1}+\frac{d}{R_2}\right)+\ldots\right\},
\end{split}\end{equation}
\begin{equation}\begin{split}
\int_{\Sigma} d^{D-1}\mathbf{x}_{\perp} \frac{\nabla H_1\cdot\nabla H_1}{H^D}=&2H\int_{0}^{R_1}dx\frac{1}{\left(L-\sqrt{R_1^2-x^2}-\sqrt{R_2^2-x^2}\right)^D}\frac{x^2}{R_1^2-x^2}\\=& \frac{H}{d^{D-\frac{1}{2}}}\sqrt{ \frac{2\pi R_1R_2}{R_1+R_2} }\frac{\Gamma\left(D-\frac{1}{2}\right)}{\Gamma(D)}\frac{2}{2D-3}\left( \frac{d}{R_1}- \frac{d}{R_1+R_2}+\ldots\right),
\end{split}\end{equation}
\begin{equation}\begin{split}
\int_{\Sigma} d^{D-1}\mathbf{x}_{\perp} \frac{\nabla H_2\cdot\nabla H_2}{H^D}=&2H\int_{0}^{R_1}dx\frac{1}{\left(L-\sqrt{R_1^2-x^2}-\sqrt{R_2^2-x^2}\right)^D}\frac{x^2}{R_2^2-x^2}\\
=& \frac{H}{d^{D-\frac{1}{2}}}\sqrt{ \frac{2\pi R_1R_2}{R_1+R_2} }\frac{\Gamma\left(D-\frac{1}{2}\right)}{\Gamma(D)}\frac{2}{2D-3}\left( \frac{d}{R_2}- \frac{d}{R_1+R_2}+\ldots\right),
\end{split}\end{equation}
\begin{equation}\begin{split}
\int_{\Sigma} d^{D-1}\mathbf{x}_{\perp} \frac{\nabla H_1\cdot\nabla H_2}{H^D}=&-2H\int_{0}^{R_1}dx\frac{1}{\left(L-\sqrt{R_1^2-x^2}-\sqrt{R_2^2-x^2}\right)^D}\frac{x^2}{\sqrt{R_1^2-x^2}\sqrt{R_2^2-x^2}}\\
=&-\frac{H}{d^{D-\frac{1}{2}}}\sqrt{ \frac{2\pi R_1R_2}{R_1+R_2} }\frac{\Gamma\left(D-\frac{1}{2}\right)}{\Gamma(D)}\frac{2}{2D-3}\left(\frac{d}{R_1+R_2}+\ldots\right).
\end{split}\end{equation}
Substituting into \eqref{eq3_27_5}, we find that  for the interaction of two cylinders,
\begin{equation}\label{eq5_18_3}\begin{split}
E_{\text{Cas}}^{\text{DE}, \text{XY}}=&b_0^{\text{XY}}\frac{H}{d^{D-\frac{1}{2}}}\sqrt{ \frac{2\pi R_1R_2}{R_1+R_2} }\frac{\Gamma\left(D-\frac{1}{2}\right)}{\Gamma(D)} \left\{1+\frac{9}{4(2D-3)}\frac{d}{R_1+R_2}-\frac{3}{4(2D-3)}\left(\frac{d}{R_1}+\frac{d}{R_2}\right)
\right.\\&\left.+\frac{2\beta_1^{\text{XY}}(D)}{2D-3}\left(\frac{d}{R_1}-\frac{d}{R_1+R_2}\right)+\frac{2\beta_2^{\text{XY}}(D)}{2D-3}\left(\frac{d}{R_2}-\frac{d}{R_1+R_2}\right)
-\frac{2\beta_{\times}^{\text{XY}}(D)}{2D-3}\frac{d}{R_1+R_2}+\ldots\right\}\\
=&b_0^{\text{XY}}\frac{H}{d^{D-\frac{1}{2}}}\sqrt{ \frac{2\pi R_1R_2}{R_1+R_2} }\frac{\Gamma\left(D-\frac{1}{2}\right)}{\Gamma(D)}\left\{1+\frac{1}{2D-3}\left(\frac{9}{4}-2\beta_1^{\text{XY}}(D)-2\beta_2^{\text{XY}}(D)-2\beta_{\times}^{\text{XY}}(D)\right)
 \frac{d}{R_1+R_2}\right.\\&\left.+\frac{1}{2D-3}\left(2\beta_1^{\text{XY}}(D)-\frac{3}{4}\right)\frac{d}{R_1}+\frac{1}{2D-3}\left(2\beta_2^{\text{XY}}(D)-\frac{3}{4}\right)\frac{d}{R_2}+\ldots\right\}.
\end{split}\end{equation}
Compare to our results \eqref{eq5_18_2}, we find that the leading terms do agree, and the next-to-leading order terms give
\begin{equation}\begin{split}
&\frac{9}{4}-2\beta_1^{\text{XY}}(D)-2\beta_2^{\text{XY}}(D)-2\beta_{\times}^{\text{XY}}(D)=-\frac{4D-5}{4},\\
&2\beta_1^{\text{DD}}-\frac{3}{4}=\frac{4D-5}{12}-\frac{(D-2)(D-3)}{3D}\frac{\zeta(D-1)}{\zeta(D+1)},\quad \beta_2^{\text{DD}}(D)=\beta_1^{\text{DD}}(D),\\
&2\beta_1^{\text{DN}}-\frac{3}{4}=\frac{4D-5}{12}-\frac{(D-2)(D-3)}{3D}\frac{\zeta(D-1)}{\zeta(D+1)}\frac{2^D-4}{2^D-1},\quad \beta_2^{\text{ND}}(D)=\beta_1^{\text{DN}}(D),\\
&2\beta_1^{\text{ND}}-\frac{3}{4}=\frac{4D-5}{12}-\frac{D^2+7D-6}{3D}\frac{\zeta(D-1)}{\zeta(D+1)}\frac{2^D-4}{2^D-1},\quad \beta_2^{\text{DN}}(D)=\beta_1^{\text{ND}}(D),\\
&2\beta_1^{\text{NN}}-\frac{3}{4}=\frac{4D-5}{12}-\frac{D^2+7D-6}{3D}\frac{\zeta(D-1)}{\zeta(D+1)},\quad \beta_2^{\text{NN}}(D)=\beta_1^{\text{NN}}(D).
\end{split}\end{equation}From these, we obtain
\begin{equation}\begin{split}
\beta_{\times}^{\text{XY}}(D)=\frac{D+1}{2}-\beta_1^{\text{XY}}(D)-\beta_2^{\text{XY}}(D),
\end{split}\end{equation}and
\begin{equation}\label{eq5_19_1}\begin{split}
\beta_1^{\text{DD}}(D)=\beta_2^{\text{DD}}(D)=&\frac{D+1}{6}-\frac{(D-2)(D-3)}{6D}\frac{\zeta(D-1)}{\zeta(D+1)},\\
\beta_1^{\text{DN}}(D)=\beta_2^{\text{ND}}(D)=&\frac{D+1}{6}-\frac{(D-2)(D-3)}{6D}\frac{\zeta(D-1)}{\zeta(D+1)}\frac{2^D-4}{2^D-1},\\
\beta_1^{\text{ND}}(D)=\beta_2^{\text{DN}}(D)=&\frac{D+1}{6}-\frac{D^2+7D-6}{6D}\frac{\zeta(D-1)}{\zeta(D+1)}\frac{2^D-4}{2^D-1},\\
\beta_1^{\text{NN}}(D)=\beta_2^{\text{NN}}(D)=&\frac{D+1}{6}-\frac{D^2+7D-6}{6D}\frac{\zeta(D-1)}{\zeta(D+1)}.
\end{split}\end{equation}When $D=3$, these agree with the results obtained in \cite{53}.

Now let us compare these results to  the results of two spheres we obtained in \cite{42}. In this case, we take the height profiles of the spheres with radii $R_1$ and $R_2$ to be
$$H_1=L_1-\sqrt{R_1^2-x_{\perp}^2}, \hspace{1cm}H_2=-L_2+\sqrt{R_2^2-x_{\perp}^2},$$where $x_{\perp}=\sqrt{x_2^2+\ldots+x_D^2}$. $L_1+L_2=L$ is the distance between the centers of the spheres.

Now, assuming $R_1<R_2$, some tedious computations give
\begin{equation}\begin{split}
\int_{\Sigma} d^{D-1}\mathbf{x}_{\perp} \frac{1}{H^D}=&\frac{2\pi^{\frac{D-1}{2}}}{\Gamma\left(\frac{D-1}{2}\right)}\int_{0}^{R_1}dx_{\perp} x_{\perp}^{D-2}\frac{1}{\left(L-\sqrt{R_1^2-x_{\perp}^2}-\sqrt{R_2^2-x_{\perp}^2}\right)^D}\\
=&\frac{\pi^{\frac{D}{2}}}{2^{\frac{D-1}{2}}\Gamma\left(\frac{D}{2}\right)d^{\frac{D+1}{2}}}\left( \frac{R_1R_2}{R_1+R_2} \right)^{\frac{D-1}{2}}
\left\{1+\frac{3(D+1)}{4}\frac{d}{R_1+R_2}-\frac{D+1 }{4 }\left(\frac{d}{R_1}+\frac{d}{R_2}\right)+\ldots\right\},
\end{split}\end{equation}
\begin{equation}\begin{split}
\int_{\Sigma} d^{D-1}\mathbf{x}_{\perp} \frac{\nabla H_1\cdot\nabla H_1}{H^D}=&\frac{2\pi^{\frac{D-1}{2}}}{\Gamma\left(\frac{D-1}{2}\right)}\int_{0}^{R_1}dx_{\perp} x_{\perp}^{D-2}\frac{1}{\left(L-\sqrt{R_1^2-x_{\perp}^2}-\sqrt{R_2^2-x_{\perp}^2}\right)^D}\frac{x_{\perp}^2}{R_1^2-x_{\perp}^2}\\
=&\frac{\pi^{\frac{D}{2}}}{2^{\frac{D-1}{2}}\Gamma\left(\frac{D}{2}\right)d^{\frac{D+1}{2}}}\left( \frac{R_1R_2}{R_1+R_2} \right)^{\frac{D-1}{2}}\left(\frac{2d}{R_1}-\frac{2d}{R_1+R_2}+\ldots\right),
\end{split}\end{equation}
\begin{equation}\begin{split}
\int_{\Sigma} d^{D-1}\mathbf{x}_{\perp} \frac{\nabla H_2\cdot\nabla H_2}{H^D}=&\frac{2\pi^{\frac{D-1}{2}}}{\Gamma\left(\frac{D-1}{2}\right)}\int_{0}^{R_1}dx_{\perp} x_{\perp}^{D-2}\frac{1}{\left(L-\sqrt{R_1^2-x_{\perp}^2}-\sqrt{R_2^2-x_{\perp}^2}\right)^D}\frac{x_{\perp}^2}{R_2^2-x_{\perp}^2}\\
=&\frac{\pi^{\frac{D}{2}}}{2^{\frac{D-1}{2}}\Gamma\left(\frac{D}{2}\right)d^{\frac{D+1}{2}}}\left( \frac{R_1R_2}{R_1+R_2} \right)^{\frac{D-1}{2}}\left(\frac{2d}{R_2}-\frac{2d}{R_1+R_2}+\ldots\right),
\end{split}\end{equation}
\begin{equation}\begin{split}
\int_{\Sigma} d^{D-1}\mathbf{x}_{\perp} \frac{\nabla H_1\cdot\nabla H_2}{H^D}=&-\frac{2\pi^{\frac{D-1}{2}}}{\Gamma\left(\frac{D-1}{2}\right)}\int_{0}^{R_1}dx_{\perp} x_{\perp}^{D-2}\frac{1}{\left(L-\sqrt{R_1^2-x_{\perp}^2}-\sqrt{R_2^2-x_{\perp}^2}\right)^D}\frac{x_{\perp}^2}{\sqrt{R_1^2-x_{\perp}^2}\sqrt{R_2^2-x^2}}\\
=&-\frac{\pi^{\frac{D}{2}}}{2^{\frac{D-1}{2}}\Gamma\left(\frac{D}{2}\right)d^{\frac{D+1}{2}}}\left( \frac{R_1R_2}{R_1+R_2} \right)^{\frac{D-1}{2}}\left( \frac{2d}{R_1+R_2}+\ldots\right).
\end{split}\end{equation}
Hence, for the interaction of two spheres, our ansatz \eqref{eq3_27_5} gives
\begin{equation}\begin{split}
E_{\text{Cas}}^{\text{DE}, \text{XY}}=&b_0^{\text{XY}}\frac{\pi^{\frac{D}{2}}}{2^{\frac{D-1}{2}}\Gamma\left(\frac{D}{2}\right)d^{\frac{D+1}{2}}} \left\{1+\frac{3(D+1)}{4}\frac{d}{R_1+R_2}-\frac{D+1}{4 }\left(\frac{d}{R_1}+\frac{d}{R_2}\right)
\right.\\&\left.+ 2\beta_1^{\text{XY}}(D) \left(\frac{d}{R_1}-\frac{d}{R_1+R_2}\right)+ 2\beta_2^{\text{XY}}(D) \left(\frac{d}{R_2}-\frac{d}{R_1+R_2}\right)
- 2\beta_{\times}^{\text{XY}}(D) \frac{d}{R_1+R_2}+\ldots\right\}\\
=&b_0^{\text{XY}}\frac{\pi^{\frac{D}{2}}}{2^{\frac{D-1}{2}}\Gamma\left(\frac{D}{2}\right)d^{\frac{D+1}{2}}}  \left\{1+ \left(\frac{3(D+1)}{4}-2\beta_1^{\text{XY}}(D)-2\beta_2^{\text{XY}}(D)-2\beta_{\times}^{\text{XY}}(D)\right)
 \frac{d}{R_1+R_2}\right.\\&\left.+ \left(2\beta_1^{\text{XY}}(D)-\frac{D+1}{4}\right)\frac{d}{R_1}+ \left(2\beta_2^{\text{XY}}(D)-\frac{D+1}{4}\right)\frac{d}{R_2}+\ldots\right\}.
\end{split}\end{equation}
With the values of $\beta_1$ and $\beta_2$ given by \eqref{eq5_19_1}, this agrees perfectly with the result we obtained in \cite{42} for two spheres when $D\neq 4$.

Now some explanations are in order. The derivative expansion technique is a formal and   non-rigorous method to obtain the small separation asymptotic behavior of the Casimir interaction energy. The result might not be correct due to some un-observed singularities in the formal derivation.   Therefore, the ansatz \eqref{eq3_27_5} can only be used as a reference for the small separation asymptotic expansion of the Casimir interaction energy, but it needs to be checked against actual computations.

\section{Conclusion}

In this work, we have considered the Casimir interaction in $(D+1)$-dimensional spacetime due to the vacuum fluctuations of massless scalar fields between a cylinder and a plate, between two parallel cylinders where one is inside the other, and between two parallel cylinders exterior to each other. We derive the explicit integral representations for the Casimir interaction energies and use them to study the large separation and small separation asymptotic behaviors of the Casimir interactions. The large separation asymptotic behaviors are easy to compute and the order of decay is smallest in the Dirichlet-Dirichlet case, and largest in the Neumann-Neumann case. The computations of the small separation asymptotic behaviors are more complicated. The leading terms are found to agree with the proximity force approximation. The results on the next-to-leading order terms are important and they exhibit some universal behaviors. In particular, we find that for the cylinder-plate case, the ratio of the next-to-leading order term to the leading order term is inversely proportional to $D$. For the case where one cylinder is inside the other, the ratio of the next-to-leading order term to the leading order term approaches the limiting value $1/2$ when $D$ is large. For the case where the two cylinders are outside each other, the  ratio of the next-to-leading order term to the leading order term approaches the limiting value $-1/2$ when $D$ is large.
Hence, we find that the ratio is bounded in  dimensions for all cases we consider. Therefore, the corrections to the proximity force approximation will not gets larger in higher dimensions, in contrast to the sphere-plate and sphere-sphere interactions, where it is found that  the ratio of the next-to-leading order term to the leading order term is proportional to $D$ when $D$ is large \cite{41,42}.

An interesting thing to note  is that our small separation asymptotic expansion for the case of Dirichlet-Dirichlet cylinder-plate interaction agrees with the result derived using derivative expansion in \cite{52}. Generalizing the $D=3$ case in \cite{53}, we postulate a general form of the derivative expansion for small separation asymptotic expansion of the scalar  Casimir interaction energy in $(D+1)$-dimensional Minkowski spacetime, for two curved surfaces with combinations of Dirichlet and Neumann boundary conditions, based on our results on the cylinder-cylinder interaction. We also check our postulate with the results we obtained for the sphere-sphere interaction in \cite{42} and find that the postulate gives correct expansion except when $D=4$.

\begin{acknowledgments}\noindent
  This work is supported by the Ministry of Higher Education of Malaysia  under   FRGS grant FRGS/1/2013/ST02/UNIM/02/2.
\end{acknowledgments}

\appendix
\section{Tabulation of constants}\label{a1}
\begin{table} [h]\caption{\label{t1}The values of $\varkappa^{\text{XY}}$ for $3\leq D\leq 6$.}

\begin{tabular}{||c|c|c|c|c|c|c|c|c|| }
\hline\hline
\multirow{2}{*}{$D$}& \multicolumn{2}{c|}{$\varkappa^{\text{DD}}$ }& \multicolumn{2}{c|}{$\varkappa^{\text{DN}}$} &\multicolumn{2}{c|}{$\varkappa^{\text{ND}}$} &\multicolumn{2}{c||}{$\varkappa^{\text{NN}}$ }  \\
\cline{2-9}
& exact & numerical & exact & numerical & exact & numerical & exact & numerical \\
\hline
&&&&&&&&\\
$3$ & $\displaystyle  \frac{7}{36} $ & $\displaystyle   0.1944  $ &$\displaystyle \frac{7}{36} $ &$\displaystyle    0.1944$ & $\displaystyle  \frac{7}{36}-\frac{160}{21\pi^2} $ & $\displaystyle   -0.5775 $ &$\displaystyle  \frac{7}{36} -\frac{40}{3\pi^2} $ &$\displaystyle -1.1565  $  \\
&&&&&&&&\\
4 & $\displaystyle   \frac{11}{60}-\frac{1}{30}\frac{\zeta(3)}{\zeta(5)} $ & $\displaystyle  0.1447 $ &$\displaystyle \frac{11}{60} -\frac{2}{75}\frac{\zeta(3)}{\zeta(5)} $ &$\displaystyle   0.1524$   & $\displaystyle \frac{11}{60} -\frac{38}{75}\frac{\zeta(3)}{\zeta(5)} $ & $\displaystyle -0.4040   $ &$\displaystyle \frac{11}{60}-\frac{19}{30}\frac{\zeta(3)}{\zeta(5)}$ &$\displaystyle   -0.5509$  \\
&&&&&&&&\\
5 & $\displaystyle \frac{5}{28}-\frac{3}{5\pi^2}  $ & $\displaystyle  0.1178 $ &$\displaystyle  \frac{5}{28} -\frac{84}{155\pi^2}$ &$\displaystyle 0.1237  $    & $\displaystyle \frac{5}{28}-\frac{756}{155\pi^2} $ & $\displaystyle  -0.3156     $ &$\displaystyle  \frac{5}{28}-\frac{27}{5\pi^2} $ &$\displaystyle   -0.3686 $ \\
&&&&&&&&\\
6 & $\displaystyle  \frac{19}{108} -\frac{2}{27}\frac{\zeta(5)}{\zeta(7)}$ & $\displaystyle  0.0998  $ &$\displaystyle \frac{19}{108} -\frac{40}{567}\frac{\zeta(5)}{\zeta(7)}  $ &$\displaystyle 0.1034  $   & $\displaystyle \frac{19}{108} -\frac{80}{189}\frac{\zeta(5)}{\zeta(7)} $ & $\displaystyle  -0.2594 $ &$\displaystyle   \frac{19}{108} -\frac{4}{9}\frac{\zeta(5)}{\zeta(7)}$ &$\displaystyle  -0.2811 $  \\
&&&&&&&&\\
\hline
\hline
\end{tabular}

\end{table}


\begin{thebibliography}{10}
\bibitem{1}
H.~B.~G. Casimir,  Proc. Kon. Nederland. Akad. Wetensch. \textbf{B51}, 793--795  (1948).

\bibitem{2} M. Bordag, G. L. Klimchitskaya, U. Mohideen, and V. M.
Mostepanenko, \emph{Advances in The Casimir Effect} (Oxford
University Press, Oxford, 2009).

\bibitem{3} S. K. Blau and M. Visser, Nucl. Phys. B \textbf{310}, 163 (1988).






\bibitem{4} E. Elizalde,  Phys. Lett. B \textbf{516}, 143 (2001).

\bibitem{5} K.A. Milton,  Grav. Cosmol. \textbf{9}, 66 (2003).

\bibitem{6} G. Mahajan, S. Sarkar and T. Padmanabhan,  Phys. Lett. B \textbf{641}, 6 (2006).

\bibitem{7} J. Ambjørn and S. Wolfram, Ann. Phys. (N.Y.) \textbf{147}, 1
(1983).


\bibitem{8} E. Elizalde, J. Math. Phys. \textbf{35}, 3308 (1994).

\bibitem{9} C. M. Bender and K. A. Milton, Phys. Rev. D \textbf{50}, 6547 (1994).
\bibitem{10} K. A. Milton, Phys. Rev. D \textbf{55}, 4940 (1996).
\bibitem{11} G. Cognola, E. Elizalde and K. Kirsten, J. Phys. A \textbf{34}, 7311 (2001).

\bibitem{12} A. Romeo and A. A. Saharian, \textbf{63}, 105019 (2001).

\bibitem{13} A. A. Saharian, Phys. Rev. D \textbf{63}, 125007 (2001).


\bibitem{14} L. P. Teo, Phys. Rev. D \textbf{82}, 085009 (2010).

\bibitem{15} L. P. Teo, J. Math. Phys. \textbf{54}, 103505 (2013).

\bibitem{16} S. Bellucci and A. A. Saharian, Eur. Phys. J. C \textbf{74}, 3047 (2014).

\bibitem{17} H. Gies, K. Langfeld, and L. Moyaerts, J. High Energy Phys. \textbf{0306},  018 (2003).

\bibitem{18} H. Gies, K. Langfeld, and L. Moyaerts, J. Phys. A \textbf{39}, 6415  (2006).

\bibitem{19} H. Gies and K. Klingm\"uller, Phys. Rev. Lett. \textbf{96}, 220401 (2006).

\bibitem{20} H. Gies and K. Klingm\"uller, Phys. Rev. Lett. \textbf{97}, 220405 (2006).

\bibitem{21} H. Gies and K. Klingm\"uller, Phys. Rev. D \textbf{74}, 045002 (2006).

\bibitem{22} A. Lambrecht, P.A. Maia-Neto, and S. Reynaud New Journal of Physics \textbf{8}, 243 (2006).

\bibitem{23} A. Bulgac, P. Magierski and A. Wirzba,   Phys. Rev. D \textbf{73}, 025007 (2006).





\bibitem{24} T. Emig, R. L. Jaffe, M. Kadar and A. Scardicchio,  Phys. Rev. Lett. \textbf{96}, 080403 (2006).


\bibitem{25} S. J. Rahi, T. Emig, R. L. Jaffe and M. Kardar, Phys. Rev. A \textbf{78}, 012104 (2008).

\bibitem{26} T. Emig,  N. Graham,  R. L. Jaffe   and M. Kardar, Phys. Rev. Lett. \textbf{99}, 170403 (2007).

\bibitem{27} T. Emig, N. Graham, R. L. Jaffe and M. Kadar, Phys. Rev. D \textbf{77}, 025005 (2008).

\bibitem{28} T. Emig and R. L. Jaffe,  J. Phys. A: Math. Theor. \textbf{41}, 164001 (2008).

\bibitem{29} T. Emig, J. Stat. Mech. \textbf{0804}, P04007 (2008).

\bibitem{30} S. J. Rahi, T. Emig, N. Graham, R. L. Jaffe and M. Kadar, Phys, Rev. D \textbf{80}, 085021 (2009).


\bibitem{31} M. Bordag,   Phys. Rev. D \textbf{73}, 125018 (2006).

\bibitem{32} M. Bordag, Phys. Rev. D \textbf{75}, 065003 (2007).




\bibitem{33} O. Kenneth and I. Klich, Phys. Rev. Lett. \textbf{97}, 160401 (2006).


\bibitem{34} O. Kenneth and I. Klich,  Phys. Rev. B \textbf{78}, 014103 (2008).


\bibitem{35} K. A. Milton and J. Wagner,  J. Phys. A: Math. Theor. \textbf{41}, 155402 (2008).

\bibitem{36} K. Milton and J. Wagner, Phys. Rev. D \textbf{77}, 045005 (2008).


\bibitem{37} D. A. R. Dalvit, F. C. Lombardo, F. D. Mazzitelli and R. Onofrio, Phys. Rev. A \textbf{74}, 020101(R) (2006).



\bibitem{38} F. D. Mazzitelli, D. A. R. Dalvit and F. C. Lombardo, New. J. Phys. \textbf{8}, 240 (2006).

\bibitem{39} F. C. Lombardo, F. D. Mazzitelli, P. I. Villar and D. A. R. Dalvit, Phys. Rev. A \textbf{82}, 042509 (2010).

\bibitem{40} L. P. Teo, Int. J. Mod. Phys. A \textbf{27}, 1230021 (2012).

\bibitem{41} L. P. Teo,  J. Math. Phys. \textbf{55}, 043508 (2014).

\bibitem{42} L. P. Teo, JHEP \textbf{2014 (05)}, 016 (2014).


\bibitem{43} L. P. Teo, Phys. Rev. D. \textbf{89}, 105033 (2014).

\bibitem{44} L. P. Teo, Phys. Rev. D \textbf{90}, 045012  (2014).

\bibitem{45} M. Bordag and V. Nikolaev,  J. Phys. A: Math. Theor. \textbf{41}, 164002 (2008).
\bibitem{46} M. Bordag and V. Nikolaev,   Phys. Rev. D \textbf{81}, 065011 (2010).
\bibitem{47} L. P. Teo, M. Bordag and V. Nikolaev, Phys. Rev. D \textbf{84}, 125037 (2011).


\bibitem{48} L. P. Teo, Phys. Rev. D \textbf{84}, 025022 (2011).

\bibitem{49} L. P. Teo, Phys. Rev. D \textbf{84}, 065027 (2011).
\bibitem{50} L. P. Teo, Phys. Rev. D \textbf{85}, 045027 (2012).
\bibitem{51} L. P. Teo, Phys. Rev. D \textbf{88}, 045019 (2013).


\bibitem{52} C. D. Fosco, F. C. Lombardo and F. D. Mazzitelli, Phys. Rev. D \textbf{86}, 045021 (2012).

\bibitem{53} G. Bimonte, T. Emig, R. L. Jaffe, and M. Kadar, Europhys. Lett. \textbf{97}, 50001 (2012).
\bibitem{54} C. D. Fosco, F. C. Lombardo and F. D. Mazzitelli, Phys. Rev. D \textbf{84}, 105031 (2011).

\bibitem{55} C. D. Fosco, F. C. Lombardo and F. D. Mazzitelli, Phys. Rev. A \textbf{89}, 062120 (2014).
\end{thebibliography}
\end{document}